# $H_2O$ and $CO_2$ Sorption in Ion-Exchange Sorbents: Distinct Interactions in Amine Versus Quaternary Ammonium Materials

Golnaz Najaf Tomaraei, Sierra Binney, Ryan Stratton, Houlong Zhuang, and Jennifer L. Wade*

**ABSTRACT:** This study investigates the $H_2O$ and $CO_2$ sorption behavior of two chemically distinct polystyrene-divinylbenzene-based ion exchange sorbents: a primary amine and a permanently charged strong base quaternary ammonium ($QA^+$) group with (bi)carbonate counter anions. We compare their distinct interactions with $H_2O$ and $CO_2$ through simultaneous thermal gravimetric, calorimetric, gas analysis, and molecular modeling approaches to evaluate their performance for dilute $CO_2$ separations like direct air capture. Thermal and hybrid (heat + low-temperature hydration) desorption experiments demonstrate that the $QA^+$-based sorbent binds both water and $CO_2$ more strongly than the amine counterparts but undergoes degradation at moderate temperatures, limiting its compatibility with thermal swing regeneration. However, a low-temperature moisture-driven regeneration pathway is uniquely effective for the $QA^+$-based sorbent. To inform the energetics of a moisture-based $CO_2$ separation (i.e., a moisture swing), we compare calorimetric water sorption enthalpies to Clausius−Clapeyron-derived total isosteric enthalpies. To our knowledge, this includes the first direct calorimetric measurement of water sorption enthalpy in a $QA^+$-based sorbent. Both methods reveal monolayer-multilayer sorption behavior for both sorbents, with the $QA^+$-based material having slightly higher water sorption enthalpies at the initially occupied strongest sorption sites. Molecular modeling supports this observation, showing higher water sorption energies and denser charge distributions in the $QA^+$-based sorbent at $\lambda_{H_2O}$ = 1 mmol/mmol$_{site}$. Mixed gas experiments in the $QA^+$-based sorbent show that not only does water influence $CO_2$ binding, but $CO_2$ influences water uptake through counterion-dependent hydration states, and that moisture swing responsiveness in this material causes hydration-induced $CO_2$ release and drying-induced $CO_2$ uptake, an important feature for low-energy $CO_2$ separation under ambient conditions. Overall, the two classes of sorbents offer distinct pathways for the $CO_2$ separation.

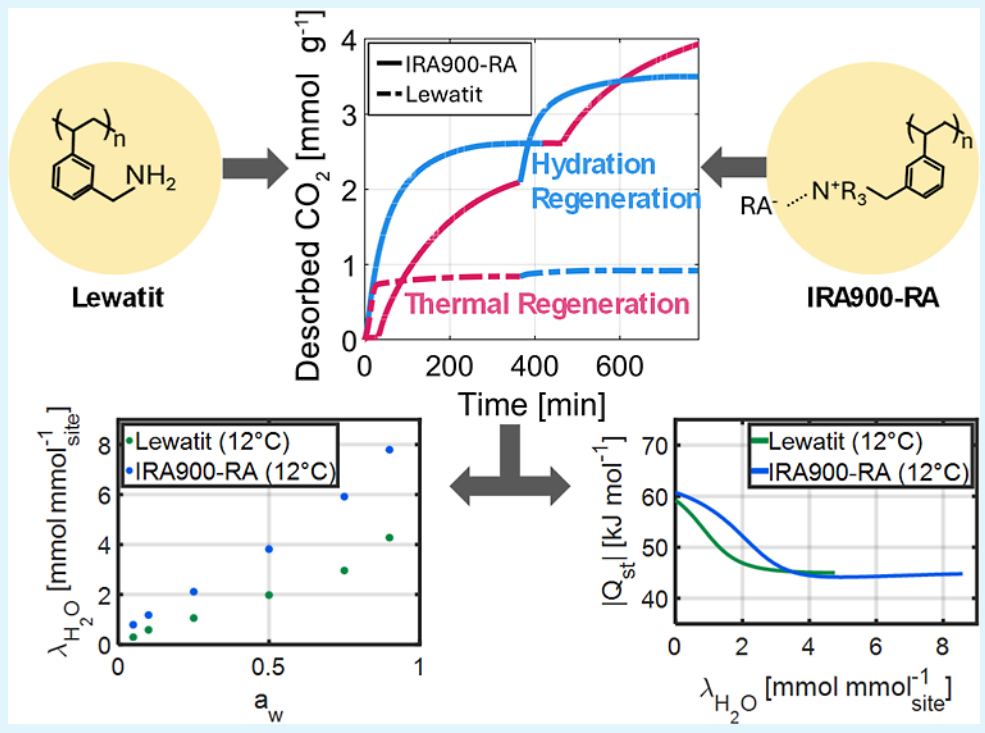

**KEYWORDS:** *direct air capture, $CO_2$ separation, primary amine, quaternary ammonium, chemisorption, sorption isotherm, Clausius−Clapeyron, GAB model, calorimetry, thermal gravimetric analysis*

## 1. INTRODUCTION

Water vapor activity plays an important role in carbon dioxide ($CO_2$) sorbents designed for dilute $CO_2$ separations like direct air capture (DAC).[1−7] For example, the $CO_2$ sorption capacity of physisorbents like zeolites is often negatively affected by competitive water sorption. Amine-based chemisorbents exhibit enhanced $CO_2$ capacity under humid conditions due to cooperative interactions between water and $CO_2$ that increase the N−$CO_2$ binding stoichiometry, though at higher water activity (>50%), the water can block the amine site from the $CO_2$, limiting uptake. Excess water binding increases the energetics of a thermal swing separation.[1−7] In $QA^+$-based anion exchange resins, hydration of bicarbonate anions is used to facilitate $CO_2$ desorption—a moisture swing—rather than thermal energy to improve energy efficiency in $CO_2$ separations.[8−11] In this study, we examine the water (de)-sorption and $CO_2$ desorption properties in two classes of $CO_2$ chemisorbents: (i) a weak base anion exchange resin with primary amine functionality (thermal swing sorbent), Lewatit VP OC 1065,[12] and (ii) a strong base anion exchange resin with trimethyl quaternary ammonium charge balanced with $CO_2$ reactive anions (moisture swing sorbent), AmberLite IRA900.[13] Both resins are polystyrene based with divinylbenzene cross-linked structure, giving rise to mesoporosity that enhances gas diffusion into the polymer. Understanding both pure and mixed water and $CO_2$ sorption thermodynamics in these materials is critical for predicting their behavior in scaled sorption processes.

Thermodynamic, kinetic, and cyclic stability data enable estimates of the energetic, productivity, and overall economic and lifecycle costs of a dilute $CO_2$ separation. This data has been provided for commercially dominant amine-based sorption chemistry,[14−17] yet it is scarce for the newer moisture-swing-based sorbents.[11] Developing robust experimental and computational approaches to quantify these mixed properties is essential for understanding $H_2O—CO_2$ cosorption interactions. In this study, we perform initial $CO_2/H_2O$ desorption using both heat and moisture exposure. This allows us to evaluate the relative effectiveness of thermal swing and moisture swing regenerations across these two classes of sorbents. The desorption study also led to a rigorous desorption protocol to ensure a consistent, desorbed state prior to pure water sorption measurements. We then validate a direct calorimetric and sorption measurement against known water sorption enthalpy values for Lewatit, extend our methodology to a moisture-swing sorbent, IRA900, and complement our findings with calculated enthalpy values using the Clausius−Clapeyron method. This integrated approach enhances our understanding of the role of water activity in $CO_2$ sorption and provides a framework for future investigations into $H_2O−CO_2$ cosorption dynamics in DAC materials.

**1.1. Sorbent Chemistries.** The amine and counteranions determine how each material interacts with $CO_2$ and $H_2O$. Under dry conditions, the primary amine groups in Lewatit VP OC 1065, henceforth called Lewatit, chemisorb $CO_2$ through alkylammonium carbamate formation, following a 2:1 $R−NH_2:CO_2$ stoichiometry (eq 1).[18−21] Under wet conditions, enhanced $CO_2$ uptake is observed in amine-based chemisorbents, potentially through the formation of alkylammonium bicarbonate or water-stabilized carbamic acid with a theoretical 1:1 $R−NH_2:CO_2$ stoichiometry at equilibrium[18−21] (eq 2). However, carbamate formation is kinetically favored and occurs much faster than bicarbonate formation. Therefore, the actual stoichiometry depends on exposure time.[19]

$$2(RNH_2) + CO_2 \leftrightarrow RNHCO_2^-RNH_3^+ \quad (1)$$

$$RNH_2 + CO_2 + H_2O \leftrightarrow RNH_3^+HCO_3^- \quad (2)$$

In contrast, the $QA^+$ groups in AmberLite IRA900, henceforth called IRA900, do not directly react with $CO_2$. Instead, the $CO_2$ reacts with an alkaline counterion, carbonate, or hydroxide forming $HCO_3^-$.[22−27] IRA900-RA represents the class of moisture swing (MS) sorbents: strong-base anion exchange resins with fixed cationic sites, typically $QA^+$ moieties, covalently bound to a polymer backbone and counterbalanced with reactive anions. The resin's affinity for $CO_2$ is moisture-dependent, with high $CO_2$ uptake under dry condition and low $CO_2$ affinity under humid conditions.[8,22,28,29] This behavior arises from hydration effects that modulate the relative ion stabilities and result in a reversible hydrolysis reaction (eq 3).

As water vapor pressure decreases, the hydration shells around the reactive anions shrink, increasing their free energy.[23,28,30] Among these reactive anions, $CO_3^{2-}$ becomes less stable than $HCO_3^-$ and $OH^-$, resulting in hydrolysis according to eq 3.

$$CO_3^{2-}.\ m_1H_2O \leftrightarrow HCO_3^-.\ m_2H_2O + OH^-.\ m_3H_2O + (m_1 - m_2 - m_3 - 1)H_2O \quad (3)$$

The generated $OH^-$ reacts with gaseous $CO_2$ to form a second $HCO_3^-$ (eq 4).

$$CO_2 + OH^-.\ m_3H_2O \leftrightarrow HCO_3^-.\ m_2H_2O \quad (4)$$

In the above expressions, $m_i$ indicates the stoichiometric size of the anion hydration clouds, where ($m_1$-$m_2$-$m_3$-1 >1) is required to satisfy Le Chatelier and mass action laws. The net MS $CO_2$ capture and release process (eq 5) is reversible. Upon re-exposure to high humidity, $CO_2$ is released and $CO_3^{2-}$ is formed and assumed to be the stable anion under wet ambient conditions.[26] However, some studies indicate the formation of $OH^-$ under high water activity and low $PCO_2$ (e.g., <100 ppm).[24,31]

$$2HCO_3^-.\ m_2H_2O + (m_1 - m_2 - m_3 - 1)H_2O \leftrightarrow CO_3^{2-}.\ m_1H_2O + CO_2 \quad (5)$$

**1.2. Sorbent Thermodynamics.** This study emphasizes $H_2O$ isotherms and their dependence on $CO_2$. $N_2$ sorption is not examined as this binding is minor at the temperature and pressures encountered in dilute $CO_2$ separations (e.g., <0.01 mmol/g at $PN_2$ = 1 bar, 20 °C) resulting in $CO_2/N_2$ selectivity above 100.[16] Prior studies have described the inverse, $CO_2$ isotherms and their dependence on $H_2O$. Young et al.[16] and Stampi-Bombelli et al.[32] have proposed water-modified Toth isotherms to describe $CO_2$ binding on dry versus wet sites in Lewatit 1065, the amine sorbent. For the (bi)carbonate-exchanged anion exchange resins, Wang et al.[9] were able to best describe the $CO_2$ isotherm using a Langmuir fit, where the $CO_2$ affinity decreased with increasing humidity. More recently, Lopez-Marques et al.[31] formalized the water-dependent $CO_2$ isotherm for IRA900-RA using a water sorption power law built into the Langmuir affinity parameter, with the exponent linked to the number of water molecules released from a $CO_2$-binding event. Kaneko and Lackner[25] described the isotherm thermodynamics to the detailed chemisorption equilibrium constants occurring in the anion exchange materials (eq 3-4), though its utility in fitting experimental data has been limited.[11]

The water sorption isotherms in the glassy polymeric sorbents measured in this study follow Type II sorption behavior,[33] which can be described using either a dual mode sorption model[34−38] or multilayer sorption models such as the Guggenheim−Anderson−de Boer (GAB) model. In the dual-mode theory, sorption occurs via Langmuir-type filling of nonequilibrium micro voids and Henry's law dissolution into the polymer matrix, originally developed for gas sorption in glassy polymers at low relative pressures and in some cases extended to describe vapor sorption at higher activities or for more soluble species. Water vapor's strong polarity and hydrogen-bonding capacity can lead to multilayer sorption and clustering, and thus the GAB model is widely used to describe water sorption in polar or charged porous polymers.

The GAB model is an extension of the Brunauer−Emmett−Teller (BET) theory and incorporates more detailed thermodynamic assumptions about the energetics of different sorption layers.[16,39,40] The GAB theory assumes localized multilayer sorption without lateral interactions on identical independent adsorption sites. It expresses equilibrium water loading ($q_{H_2O}$) as a function of water activity ($a_w$) using eq 6, where $q_m$ is the monolayer capacity, and c and k are affinity parameters related to the first and subsequent multilayers, respectively. As $k$ approaches 1, multilayer water becomes indistinguishable from the bulk, and the GAB model reduces to the BET form. The model distinguishes three energetic regimes: a tightly bound monolayer, structured multilayers with intermediate binding energies between the monolayer and bulk liquid, and bulk-like layers where sorption energy equals the latent heat of condensation.[16,39,40]

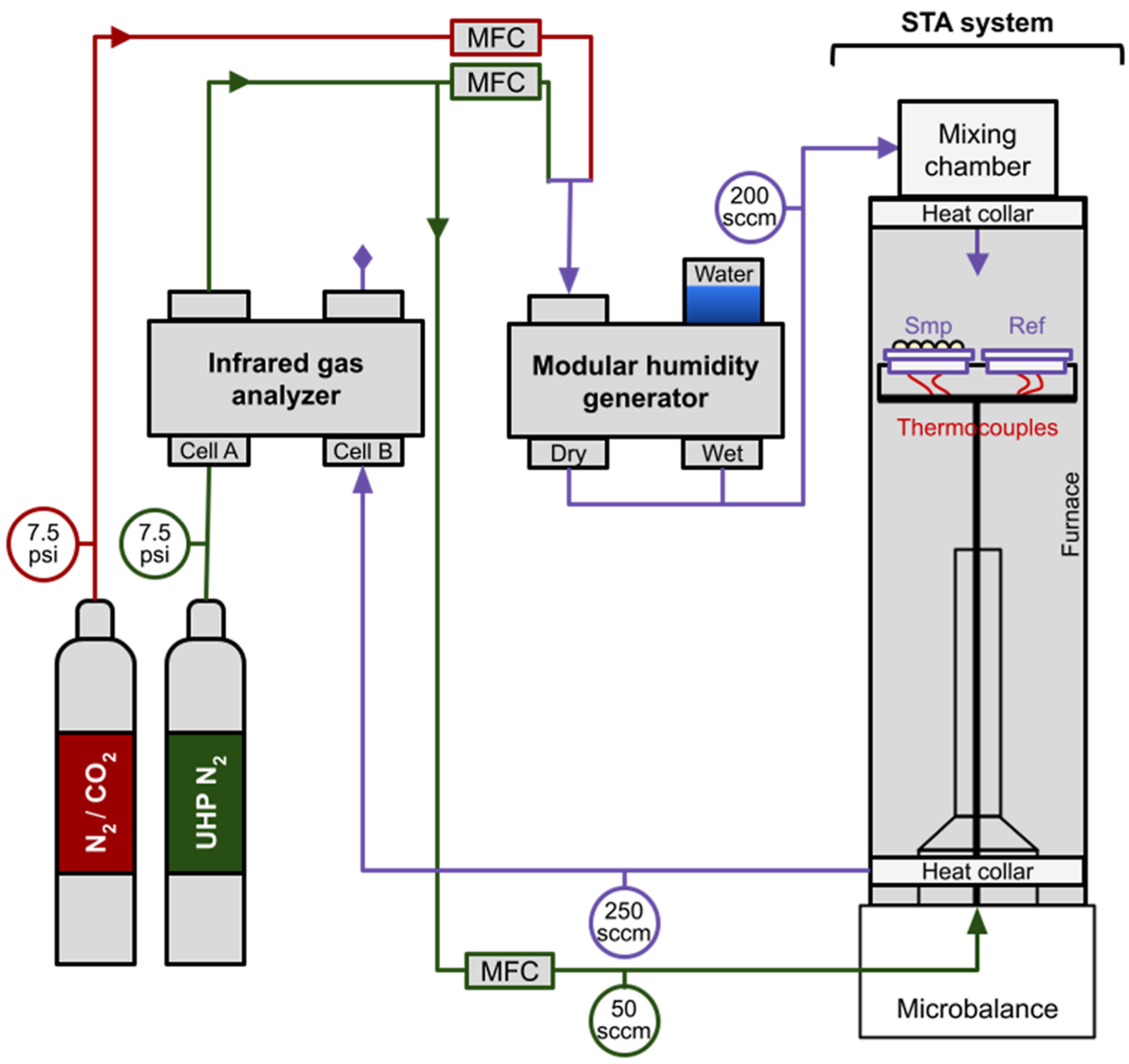

**Figure 1.** Schematic of the open-flow system used for thermalgravimetric, calorimetric, and evolved gas analyses in mixed sorption experiments.

$$q_{H_2O} = \frac{q_m k c a_w}{(1 - k a_w)[1 + (c - 1) k a_w]} \quad (6)$$

To estimate the molar net isosteric heat of sorption ($\Delta H_{is}$) from water vapor sorption isotherms measured at multiple temperatures, the Clausius−Clapeyron (CC) relation is commonly employed. This thermodynamic expression relates changes in $a_w$ with temperature at a fixed equilibrium water loading, $q_{H_2O}$, under the assumption of temperature-invariant $\Delta H_{is}$. In practice, $a_w$ as a function of $q_{H_2O}$ is derived from the GAB model using the fitted parameters. The slope of a plot of $\ln(a_w)$ versus $1/T$ at a constant $q_{H_2O}$ yields $\Delta H_{is}$ (eq 7).

$$\Delta H_{is} = -R\left(\frac{\partial \ln(a_w)}{\partial\left(\frac{1}{T}\right)}\right)_{q_{H2O}} \quad (7)$$

$\Delta H_{is}$ is defined as the difference between the total isosteric heat of sorption ($Q_{st}$) and the temperature-dependent enthalpy of vaporization of water, $\Delta H_v(T)$ (eq 8), representing the excess energy required to desorb water from the sorbent compared to bulk water evaporation.

$$Q_{st} = \Delta H_{is} + \Delta H_v(T) \quad (8)$$

## 2. MATERIALS AND METHODS

**2.1. Materials.** Lewatit VP OC 1065 resin was purchased from Sigma-Aldrich and air-dried at room temperature prior to use. Based on the supplier-reported capacity of 1.4 mequiv/mL, a wetted bed density of 1040 g/L, and a dry bed density of 630 g/L, the site density was estimated as 2.1 mmol/g dry polymer.

AmberLite IRA900 was purchased in chloride form (IRA900-Cl) from Sigma-Aldrich and ion exchanged into the bicarbonate form following the procedure established by Lopez-Marques et al.[31] One gram of IRA900-Cl was continuously stirred in 50 mL of a 0.5 M $KHCO_3$ solution in deionized (DI) water for 24 h. This process was repeated twice with a fresh $KHCO_3$ solution, and DI rinses were performed between steps. The extent of ion exchange was measured using Hach chloride strips and summed across three ion-exchange washes to determine the total ion exchange capacity, which was found to be 3.5 ± 0.1 mmol of $HCO_3^-$/g of dry polymer, with uncertainty tied to the detection limit of the chloride strips. Finally, the resin was stirred in DI water for 24 h to remove any excess salt. Although, initially in the $HCO_3^-$ form, we first desorb the intrinsic $CO_2$ through a combination of thermal desorption and low-temperature moisture-driven desorption steps (Section 2.3.1). This process leads to a shift in the reactive anion from $HCO_3^-$ to hydroxide ($OH^-$), with the carbonate anion ($CO_3^{2-}$) as a potential intermediate. Throughout this work, we refer to the bicarbonate-exchanged resin as IRA900-RA to indicate its reactive anion state, recognizing that the dominant anion species ($HCO_3^-$, $CO_3^{2-}$, or $OH^-$) will shift under different partial pressure and thermal conditions, although the anion was not directly identified. We further evaluated IRA900-Cl to act as a control by which to compare the impact of the reactive anion on water sorption and $CO_2$ desorption behavior.

**2.2. Surface Area and Pore Size Analysis.** Nitrogen physisorption measurements were conducted at 77 K using a Micromeritics TriStar II instrument to characterize the surface area and porosity of Lewatit, IRA900-Cl, and IRA900-RA. Samples were analyzed without prior degassing. Surface area was determined using the BET method, and pore size distributions and cumulative pore volumes were calculated from the desorption branch using the Barrett−Joyner−Halenda (BJH) method.[41] The desorption branch is often used with the BJH method.[42]

**2.3. Simultaneous Gravimetric, Calorimetric, and Evolved Gas Analysis.** Thermal decomposition, mixed gas desorption, and pure water sorption experiments were conducted using a Netzsch STA 449 F3 Jupiter, which combines thermogravimetric analysis (TGA) and differential scanning calorimetry (DSC) with gas analysis in an open flow-configuration. In the schematic (Figure 1), "Smp" refers to the sample within the STA chamber, where temperature, mass change, and heat flow are measured, and "Ref" refers to the reference used for baseline temperature and heat flow monitoring; both Smp and Ref are monitored by thermocouples. The STA system was coupled with a modular humidity generator system (MHG32, Prohumid) to control the relative humidity (RH) of the mixed gas composition that flows through the sample chamber. Ultrahigh purity (UHP) $N_2$ and mixed $CO_2/N_2$ gas are combined to control the partial pressure of $CO_2$ surrounding the sorbent sample at a total flow rate of 200 sccm. The mixed gas stream enters the MHG32 where it is further split into dry and wet flows, mixed at an external unit attached to the STA chamber to achieve the desired RH (Figure 1).

The STA's output stream, consisting of purge and protective gases, was directed to a LI-7000 differential, nondispersive infrared gas analyzer (IRGA) for in situ measurement of $CO_2$ (ppm) and $H_2O$ (ppt) concentrations. UHP $N_2$ flowed through Cell A of the LI-7000 (reference cell) before mixing with the $CO_2$ stream, while the STA's outlet was directed to Cell B (sample cell). $CO_2$ and $H_2O$ concentrations were calculated based on the differential infrared absorption between the two cells. The total flow entering Cell B consisted of 200 sccm of UHP $N_2$ or mixed $CO_2/N_2$ directed through the sample chamber, along with 50 sccm of protective UHP $N_2$ bypassing the sample to shield the microbalance. This protective flow did not mix at the sample interface and contributed to an effective 20% dilution of the analyte concentrations.

Real-time $CO_2$ concentrations from the gas analyzer were converted to molar flow rates (mol/s) and then integrated over time to calculate the cumulative uptake or release. These values were normalized by the sample dry mass to yield sorbent loading, q(t), in mmol $g^{-1}$. The mass change due to the $CO_2$ uptake or release was subtracted from TGA data to determine the mass change due to the $H_2O$ uptake or release. This mass-difference approach enabled measurement of bicomponent counter sorption in moisture-swing sorbents. It also provided more accurate $H_2O$ quantifications than direct gas analyzer readings.

We assumed pseudo-first-order kinetics represented by the linear driving force differential in eq 9. This expression relates the difference in loading from an equilibrium (isotherm) value, $q^*$, to an overall mass transfer coefficient, $k$, in $s^{-1}$. If a clear plateau was not reached, $q^*$ was taken as the final desorbed value at the end of the experimental segment. The integrated linear form used to extract the mass transfer coefficient using a least-squares linear regression is given by eq 10.

$$\frac{dq}{dt} = k(q^* - q(t)) \tag{9}$$

$$\ln\left(\frac{q(t) - q^*}{q(0) - q^*}\right) = -kt \tag{10}$$

Representative raw TGA and DSC signals, logged RH data from MHG32, and $CO_2$ and $H_2O$ concentrations from LI-7000 for IRA900-RA are provided in Figure S1. Measurement uncertainties were evaluated using standard error propagation methods (see the Supporting Information).

*2.3.1. $CO_2$ Desorption Experiments.* To develop effective desorption strategies for removing initially sorbed $CO_2$ and $H_2O$ from sorbents prior to water sorption experiments, we first conducted thermal decomposition tests on Lewatit, IRA900-Cl, and IRA900-RA. To normalize the starting condition, samples were exposed to 400 ppm of $CO_2$ in UHP $N_2$ for 2 h and then flushed with UHP $N_2$ for 5 min to reset the gas analyzer's baseline. The temperature was then ramped to 400 °C at 10 K/min under a UHP $N_2$ flow. These tests identified the thermal degradation onset for each sorbent. Throughout the thermal decomposition measurements, the evolved $CO_2$ and $H_2O$ gas concentrations measured by IRGA were normalized to the dry mass of each sample and reported in units of ppm/g.

Informed by these decomposition results, separate thermal desorption tests were performed by heating each sorbent to 100 °C and holding for 10 h under 200 sccm UHP $N_2$ flow. Higher temperatures were not attempted based on our high-temperature $CO_2$ desorption profiles and due to the known quaternary trimethylammonium decomposition above 120 °C.[43−45]

Building on these findings and knowing the moisture swing (MS) responsiveness of IRA900-RA, we developed and tested two desorption strategies: (1) heating the sample to 100 °C and holding under $N_2$ flow and (2) exposing the material to 95% RH at 25 °C. These strategies were tested in both sequential orders to evaluate their effect on $CO_2$ release in both IRA900-RA and Lewatit. All sorbents were pre-exposed to 400 ppm of $CO_2$ at 20% RH for 5 h to standardize initial loading. Based on these tests, a 10 h thermal desorption at 100 °C under $N_2$ was selected for Lewatit. The final desorption protocol for IRA900-RA consisted of 5 h of thermal desorption at 100 °C, followed by 5 h of exposure to 95% RH at 25 °C.

*2.3.2. Water Sorption Experiments.* For Lewatit experiments, approximately 7 mg of the air-dried sample was loaded into the STA at each temperature condition (12 °C, 25 °C, and 40 °C). The sample was first heated to 100 °C at a rate of 10 K/min and held at 100 °C for 10 h to remove presorbed $CO_2$ and $H_2O$. The sample was then cooled and maintained at the desired isothermal temperature (12 °C, 25 °C, or 40 °C) within 0.3 °C precision. Once the target temperature was reached, RH was sequentially adjusted using the MHG32, increasing stepwise from 0% to 5%, 10%, 25%, 50%, 75%, and 90%, with each RH level maintained for 5 h. The sequence was then reversed, decreasing RH stepwise back to 0% and again holding 5 h at each interval. All experimental segments were conducted under a constant flow of 200 sccm of UHP $N_2$.

For IRA900 experiments, approximately 7 mg of the sample was loaded into the STA at each temperature condition (12 °C, 25 °C, and 40 °C). After the heat + hydration desorption protocol, RH was reduced stepwise from 95% to 0%—using the levels 90%, 75%, 50%, 25%, 10%, and 5%—holding at each level for 5 h and then increased back to 90% in the same stepwise manner. This RH cycling mirrored the Lewatit procedure but in reverse order, starting with desorption and transitioning to sorption cycles.

To investigate the effect of $CO_2$ on water sorption behavior, additional experiments were conducted on IRA900-RA at 25 °C under two $CO_2$ concentrations: 400 and 3000 ppm in $N_2$. These mixed-gas experiments followed the same heat + hydration desorption protocol described in Section 2.3.1 and used the same RH cycling procedure. Negative peaks in the $CO_2$ signal during RH-decreasing steps corresponded to the $CO_2$ sorption process (bicarbonate formation, eq 5), while positive peaks during RH-increasing steps indicated $CO_2$ desorption (moisture-induced conversion of bicarbonate to carbonate, eq 5). $CO_2$ uptake/release during these cycles was analyzed using the approach previously described, where these peaks were integrated and normalized to obtain $q_{CO_2}$. The $CO_2$ mass change in each RH segment was subtracted from the total mass change in that segment, as measured by TGA, to obtain the corresponding $H_2O$ uptake/release.

The DSC signal recorded exothermic heat flow during stepwise $H_2O$ uptake (increasing RH) and endothermic heat flow during $H_2O$ release (decreasing RH). The heat flow signal was integrated over each RH transition using Netzsch Proteus software and converted from $\mu$V•s to kJ using a Ga sensitivity factor obtained by calibrating the DSC with Ga metal under similar gas flow conditions. These energy values were normalized by the number of moles of water gained or lost, as determined by TGA over the same DSC integration time interval. The directly measured calorimetric heats of sorption were compared to isosteric heats of sorption derived from the Clausius−Clapeyron analysis (eq 7).

**2.4. Molecular Simulation.** To model Lewatit, we constructed a system containing 10 monomer units confined within a cubic box with a side length of 20 Å. The resulting molecular structures are placed in a cubic supercell with a side length of approximately 25 Å. Similarly, for the IRA900-RA model, we use 8 monomers, each with a nitrogen site bonded to one bicarbonate ion. In both models, one water molecule is added to each nitrogen site to model the case of $\lambda_{H_2O} = 1$ mmol $mmol^{-1}_{site}$. Figure 2 shows the atomic structures used in the simulations of the two polymers.

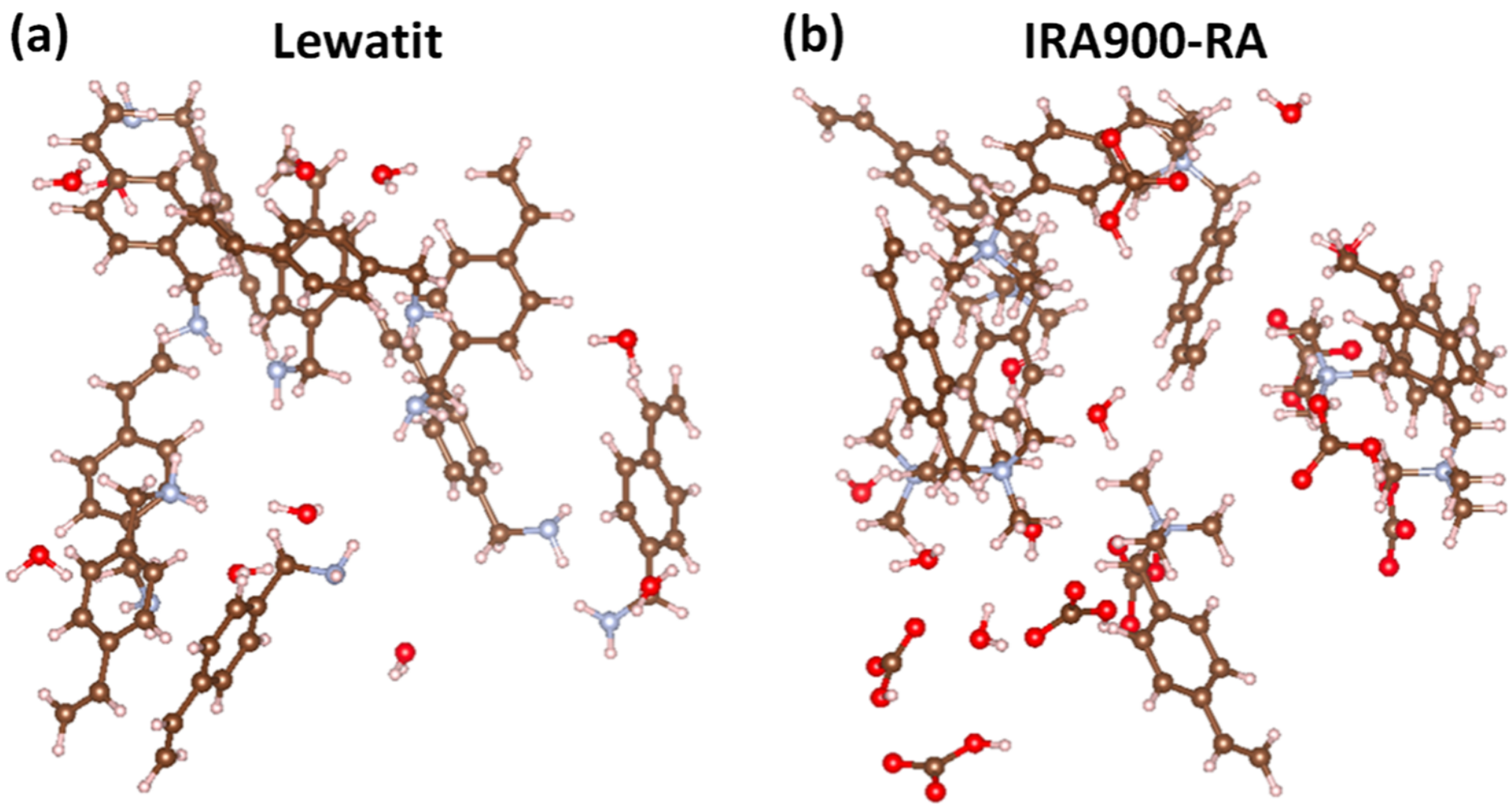

**Figure 2.** Atomic structures of (a) Lewatit and (b) IRA900-RA used in density functional theory calculations. White, brown, blue, and red spheres represent hydrogen, carbon, nitrogen, and oxygen atoms, respectively.

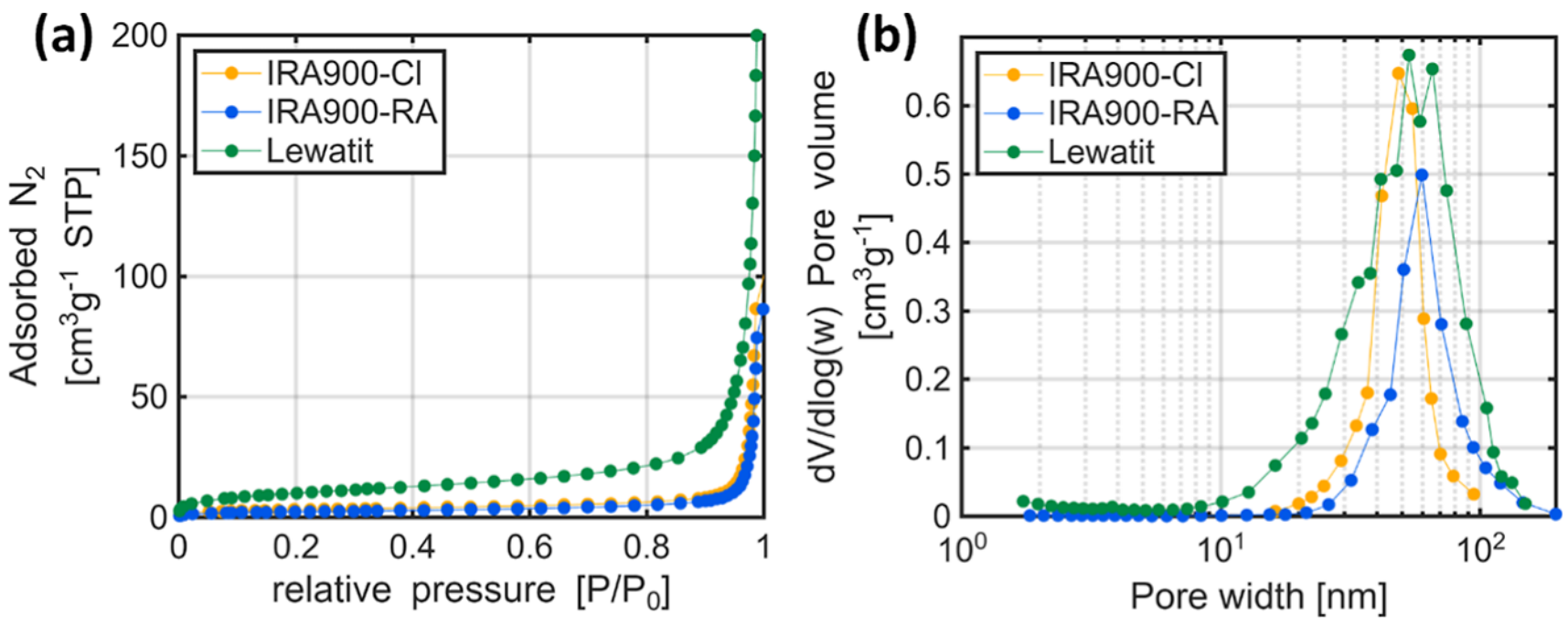

**Figure 3.** Pore structure characterization of Lewatit and IRA900 sorbents. (a) $N_2$ adsorption isotherms at 77 K, showing differences in nitrogen uptake behavior between Lewatit and IRA900 in both Cl and RA forms. IRA900 in both forms sorbs less nitrogen overall. (b) BJH pore size distributions, showing narrower distributions centered at larger pore diameters for IRA900-Cl and IRA900-RA compared to Lewatit.

We perform all density functional theory (DFT) calculations using the Vienna Ab Initio Simulation Package (VASP).[46] A plane-wave energy cutoff of 400 eV is applied. The supercell structures—including both lattice constants and atomic positions—are fully relaxed over 300 ionic steps, with the convergence criterion set such that the energy change between steps is less than 0.1 eV. Electron−electron exchange−correlation interactions are described using the Perdew−Burke−Ernzerhof (PBE) functional,[47] while electron−nucleus interactions are treated using standard projector augmented-wave (PAW) potentials.[48] All supercell calculations use a single *k*-point due to the large size of the supercell.

## 3. RESULTS AND DISCUSSION

### 3.1. BET Surface Area and BJH Pore Size Distribution.

The commercial sorbents evaluated in this study comprise similar chemical and mesoporous structures in that both materials comprise polystyrene-based backbones with divinylbenzene cross-linking. To assess the impact of the polymer textural properties on sorption behavior (e.g., mesoporosity and surface area), $N_2$ physisorption was performed to evaluate BET surface area and BJH pore size distribution (Figure 3 and Table S1). Lewatit exhibited the highest surface area (36.7 $m^2$ $g^{-1}$) and total pore volume (0.31 $cm^3$ $g^{-1}$), with an average pore width of ∼30 nm. In contrast, IRA900-Cl and IRA900-RA showed lower surface areas (11.5 and 8.3 $m^2$ $g^{-1}$, respectively) and smaller total pore volumes (0.15 and 0.13 $cm^3$ $g^{-1}$, respectively) with narrower pore size distributions centered at larger average pore widths (∼45 and ∼52 nm, respectively). This result is relevant when comparing the higher water sorption of IRA900 over Lewatit, despite the smaller surface area, discussed later in Section 3.3.

### 3.2. $CO_2$ Desorption. *3.2.1. Desorption in Pure $N_2$.*

Accurate water sorption measurements require removal of presorbed $CO_2$ and $H_2O$ from the sorbents. Thermal decomposition and desorption analysis of the sorbents identified the temperature range where $CO_2$ begins to desorb while avoiding degradation of functional groups and the polymer backbone.

The thermal decomposition experiments reveal clear differences among the three samples (Figure 4). TGA in Figure 4a shows that Lewatit undergoes gradual mass loss with a relatively high thermal stability up to 350 °C, followed by a more rapid decline. In contrast, IRA900-RA shows earlier onset of mass loss, with a steep drop occurring between 150 and 200 °C. IRA900-Cl, however, displays a more gradual mass loss, with accelerated decomposition occurring between 250 and 350 °C.

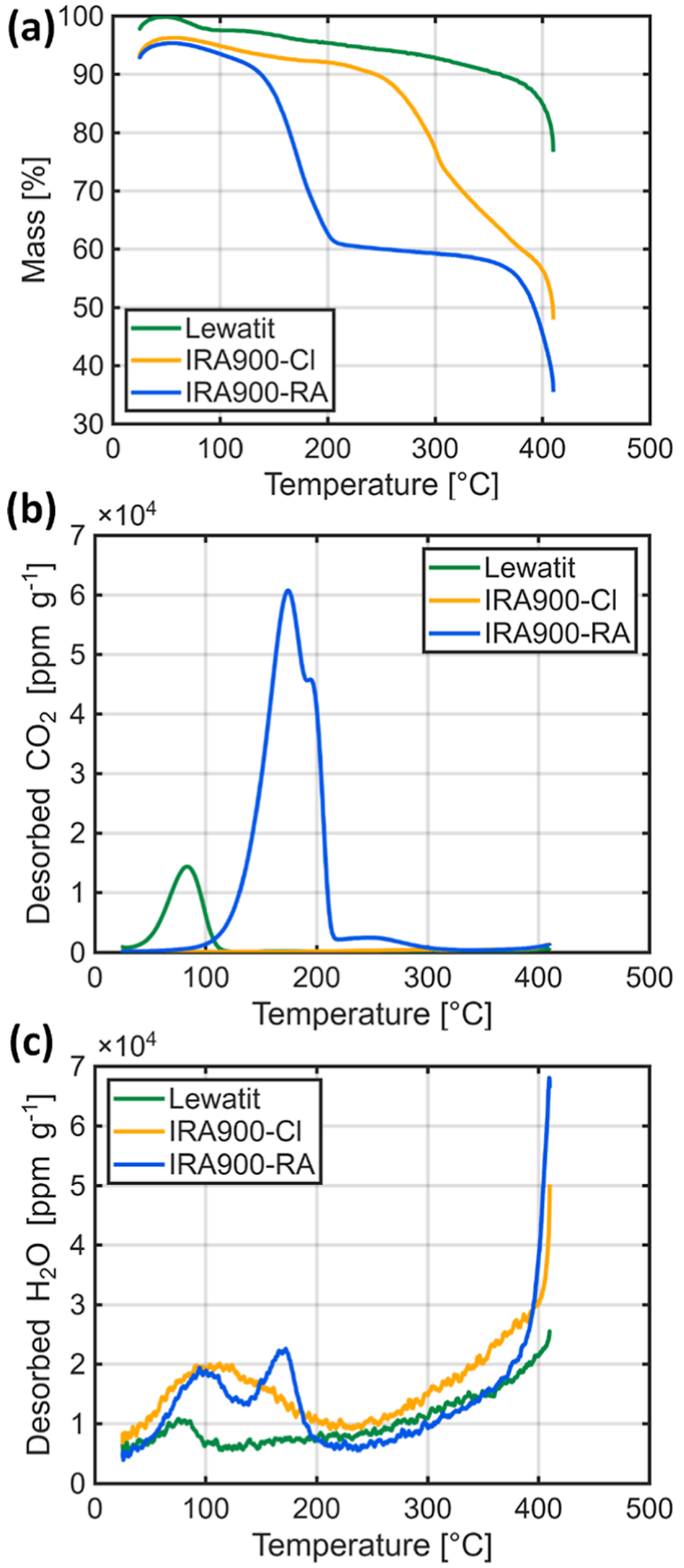

**Figure 4.** Thermal decomposition of Lewatit, IRA900-RA, and IRA900-Cl. (a) TGA mass loss profiles. Lewatit exhibits gradual mass loss up to 350 °C, whereas IRA900-RA shows earlier onset of decomposition (150−200 °C), and IRA900-Cl displays enhanced thermal stability with delayed decomposition (250−350 °C) compared to IRA900-RA. (b) $CO_2$ desorption profiles during temperature ramp under $N_2$, showing early $CO_2$ release from Lewatit (∼40 °C) and delayed release in IRA900-RA (∼100−180 °C). Minimal $CO_2$ desorption is observed in IRA900-Cl. (c) $H_2O$ desorption profiles showing earlier water release from Lewatit compared to IRA900-RA and IRA900-Cl. A secondary $H_2O$ release between 130 and 180 °C in IRA900-RA indicates decomposition-related desorption.

Lewatit shows a $CO_2$ release starting around 40 °C, with a peak near 100 °C, indicative of the thermal decomposition of carbamate, the chemisorption product between a primary amine and $CO_2$ (Figure 4b). IRA900-RA shows $CO_2$ desorption beginning around 100 °C and increasing to a peak at 180 °C. In the absence of $CO_2$ and moisture, the reactive anion in IRA900-RA decomposes from the starting $HCO_3^-$ state into a hydroxide, $OH^-$, releasing $CO_2$ (reverse of eq 4). Further, hydroxide is a well-known nucleophile that can attack the polymer backbone near the site of the $QA^+$ (e.g., Hofmann $\beta$ elimination and nucleophilic substitution[49−51]), which begins above 120 °C.[43−45] Both bicarbonate decomposition and polymer degradation due to nucleophilic attack cause the mass loss and large $CO_2$ release from IRA900, especially when comparing to the same material in the Cl-state, IRA900-Cl, which shows negligible $CO_2$ desorption throughout the temperature range (Figure 4b). This inherent overlap between bicarbonate decomposition and polymer degradation, driven by the thermal instability of $QA^+$ in the presence of hydroxide, underscores why these anion-exchange sorbents are more suitable for low-temperature $CO_2$ separations.

As shown in Figure 4c, $H_2O$ desorption in Lewatit begins around 50 °C and gradually increases to a peak near 75 °C. Both IRA900-RA and IRA900-Cl show greater $H_2O$ desorption than Lewatit, starting around 50 °C. The difference in low-temperature $H_2O$ desorption between the materials is further reflected in the water sorption isotherms discussed in Section 3.3.1. In IRA900-RA, a secondary increase in $H_2O$ desorption is observed between 130 and 180 °C, consistent with the thermal degradation of bicarbonate anions and $QA^+$ groups in the presence of reactive anions. A significant rise in $H_2O$ desorption above 350 °C in all resins is attributed to decomposition of the polymer backbone.

Decomposition experiments show that the release of $CO_2$ from IRA900-RA requires higher temperatures, indicating stronger binding of $CO_2$ to the reactive anions compared to that of $CO_2$ bound to the primary amine found in Lewatit. However, as shown by the IRA900-RA decomposition curve (Figure 4a), the $QA^+$ groups begin to thermally decompose above 120 °C, meaning that we cannot rely on the thermal desorption of $CO_2$ above this temperature without decomposing the resin. As a result, thermal regeneration of $CO_2$ must occur near 100 °C where the desorption kinetics are slow. These findings reinforce that IRA900-RA requires higher thermal energy to drive $CO_2$ desorption and thus is unsuitable for a conventional thermal swing sorption process. Instead, it is low-temperature, moisture-driven $CO_2$ desorption that is most interesting for the carbonate-based anion exchange materials.

To assess the desorption performance under these constraints, we next conducted thermal desorption experiments at 100 °C, below the degradation threshold for $QA^+$ groups. Thermal desorption (10 K/min ramp to 100 °C in UHP $N_2$, held for 10 h) revealed distinct kinetics and temperature-dependent behaviors between the materials (Figure 5). Lewatit released smaller amounts of $CO_2$ than IRA900-RA, as seen in Figure 5a. Further, Lewatit exhibited two distinct rate constants for $CO_2$ thermal desorption (Figure 5b). As a reminder, all kinetic information was analyzed using a pseudo-first-order model (eq 10). We note that the kinetic fits are not perfectly linear, indicating an interplay of complex transport phenomena, including heat transfer, since the temperature was ramped to 100 °C. Details of the rate-limiting transport phenomena were not in scope with this study. The faster process begins around 40 °C and dominates during the ramp to 100 °C, with a rate constant of 0.24

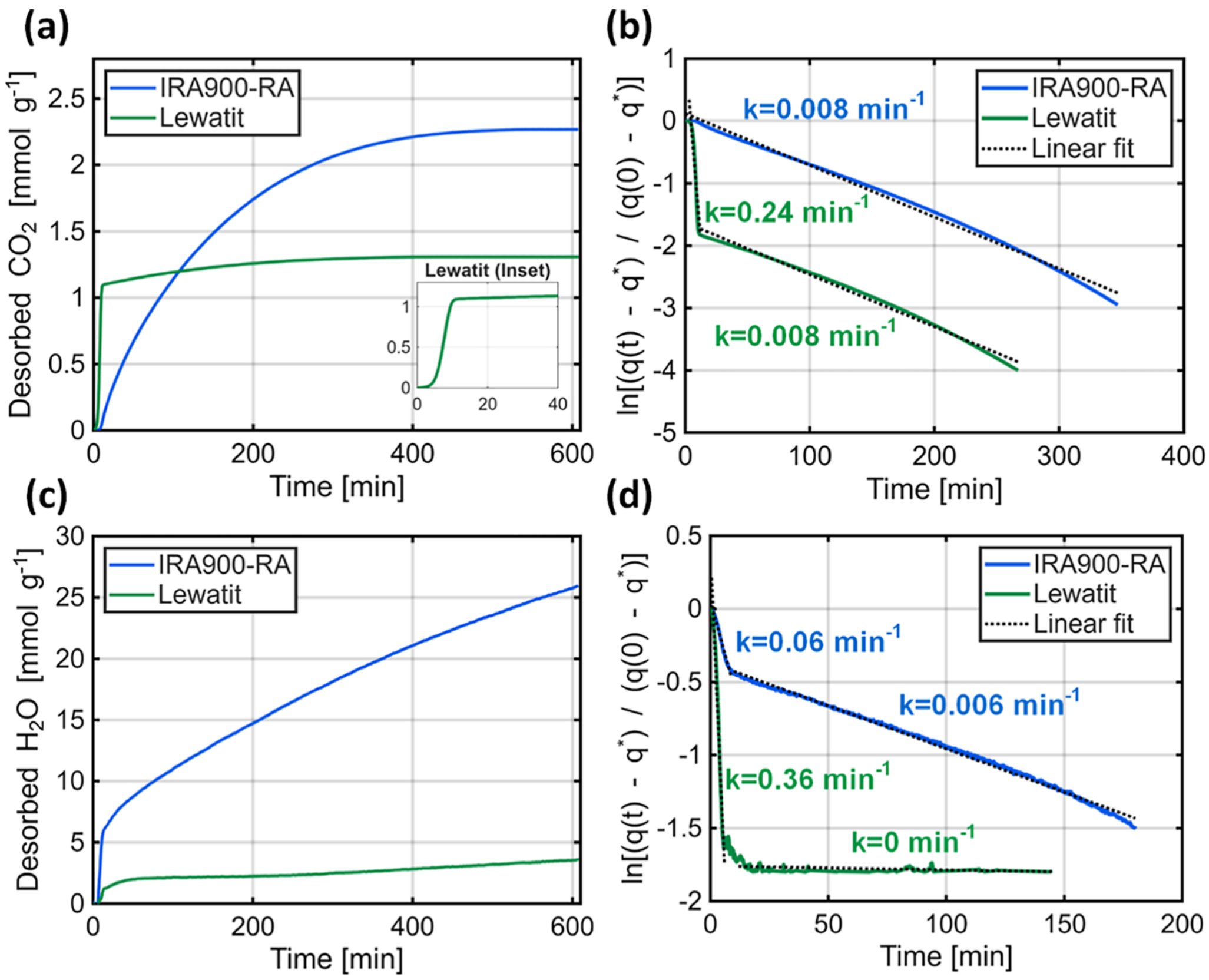

**Figure 5.** Comparison of $CO_2$ and $H_2O$ thermal desorption from IRA900-RA and Lewatit. (a) $CO_2$ desorption profiles during thermal desorption at 100 °C and (b) first-order kinetic analysis of $CO_2$ desorption showing distinct rate constants for Lewatit and IRA900-RA, with Lewatit showing faster $CO_2$ desorption kinetics during the early stage of heating. (c) $H_2O$ desorption profiles during thermal desorption at 100 °C. (d) Kinetic analysis of $H_2O$ desorption revealing two distinct desorption steps in both materials, with IRA900-RA exhibiting desorption at higher temperatures and with slower desorption kinetics, consistent with stronger water binding.

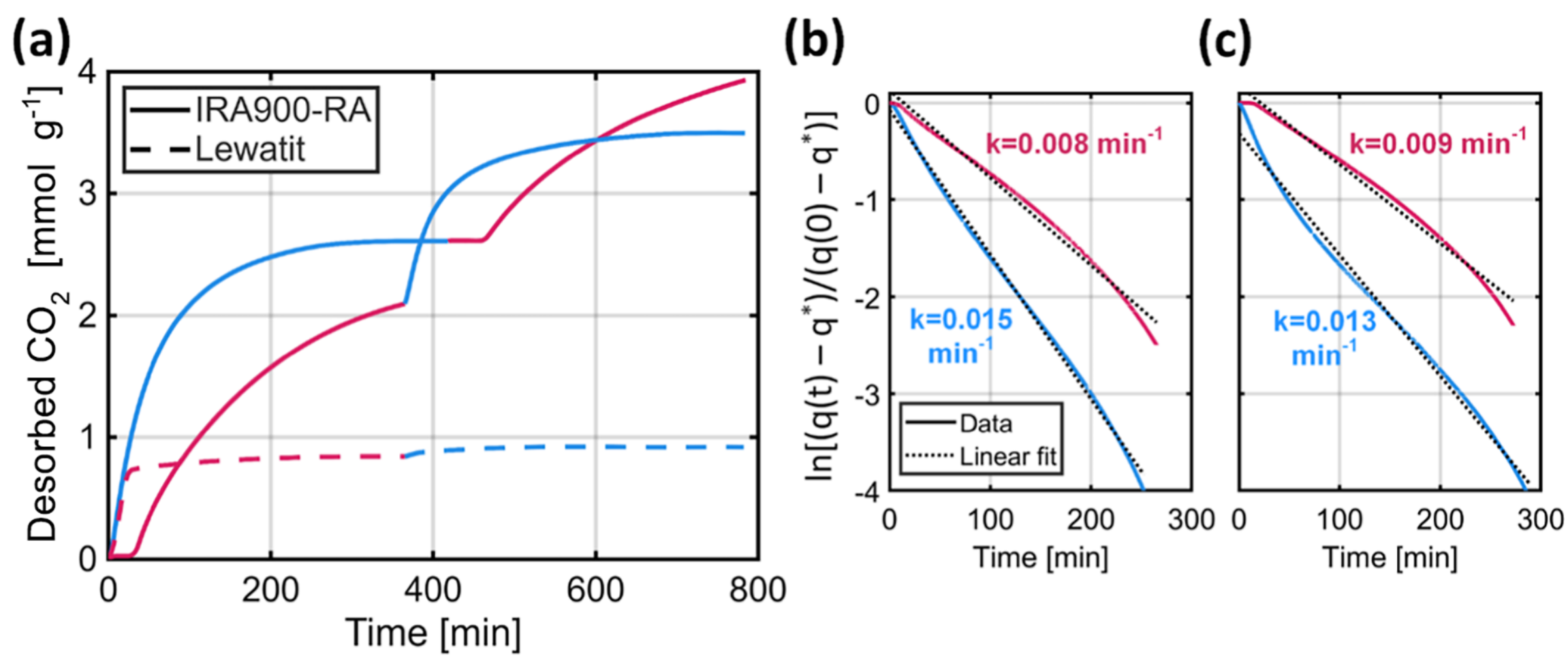

**Figure 6.** (a) $CO_2$ desorption profiles for IRA900-RA under two regeneration protocols: heat (red) followed by hydration (blue) and hydration (blue) followed by heat (red). $CO_2$ is released in both steps; moisture-driven desorption is unique to IRA900-RA. Lewatit, shown for comparison (dashed), does not exhibit moisture-driven $CO_2$ release. Pseudo-first-order kinetics of $CO_2$ desorption in IRA900-RA for (b) the first step and (c) the second step in each protocol, heat (red) or hydration (blue). The slopes of the linear fits give k ($min^{-1}$). Hydration is faster than heating, with only minor influence from the order in which the steps were applied.

$min^{-1}$—an order of magnitude faster than the slower process (0.008 $min^{-1}$). IRA900-RA, however, desorbs $CO_2$ more slowly, with nearly all release occurring during the 100 °C isothermal hold, at a rate constant of 0.008 $min^{-1}$ (Figure 5b). The higher temperature of $CO_2$ desorption observed in IRA900-RA indicates more stable free energy of $CO_2$ binding over that of Lewatit. This observation is consistent with the higher desorption onset temperature seen in the thermal decomposition experiments. Further, the thermal desorption rate of IRA900-RA matches that of the slower $CO_2$ desorption process in Lewatit suggesting that this slower process is gas diffusion limited. This difference does not indicate a fundamentally different binding mechanism, as both materials involve chemisorption. Rather, it reflects the distinct thermal stabilities of the chemisorbed $CO_2$ species. Carbamates formed on amines decompose below 100 °C, which accounts for the fast $CO_2$ release observed in Lewatit during the temperature ramp. In contrast, bicarbonates typically decompose above 100 °C. At 100 °C, IRA900-RA shows only a slow $CO_2$ release due to a slight equilibrium shift driven by the absence of $CO_2$ in the gas phase, while the main bicarbonate decomposition occurs at higher temperatures (Figure 4b).

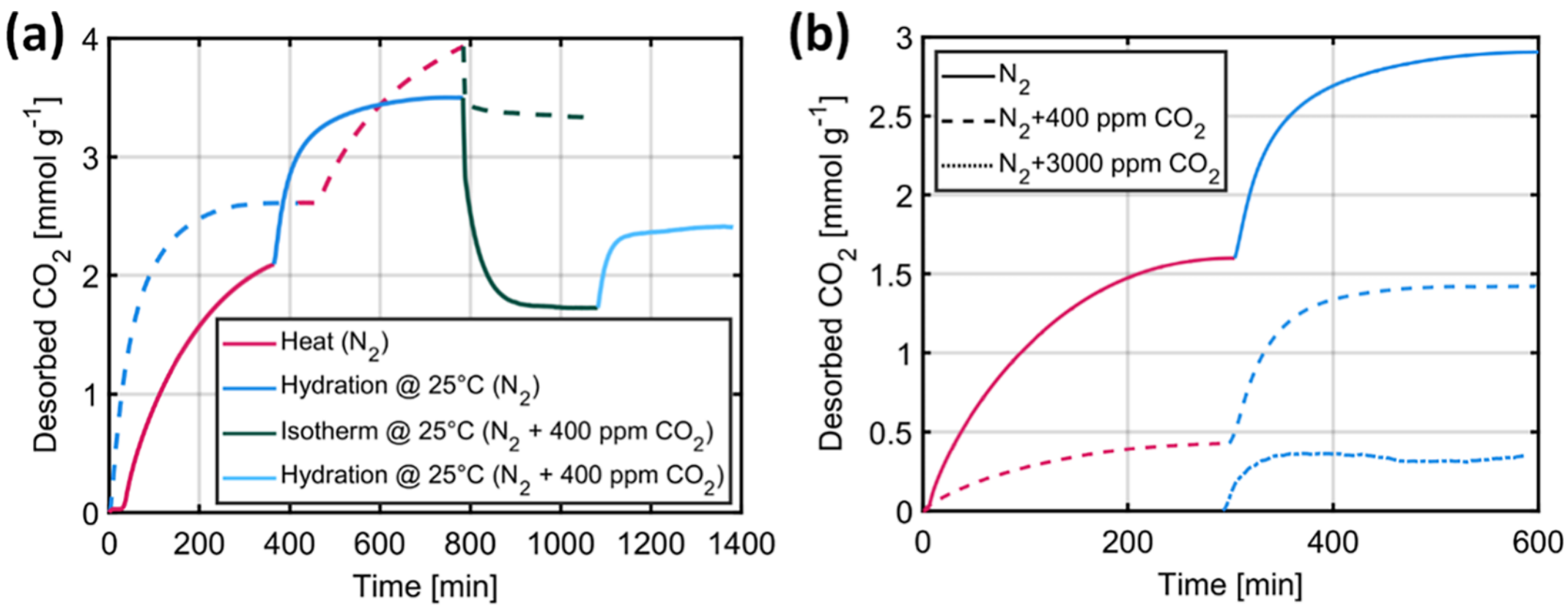

**Figure 7.** $CO_2$ desorption behavior of IRA900-RA. (a) $CO_2$ desorption and uptake following two desorption protocols: hydration followed by heating (dashed line) and heating followed by hydration (solid line), both in UHP $N_2$. These were followed by $CO_2$ uptake at 400 ppm of $CO_2$ and 0% RH for 5 h. The heat-first protocol led to greater $CO_2$ sorption and was therefore selected for subsequent water sorption studies. $CO_2$ desorption during 95% RH exposure in 400 ppm of $CO_2$ confirms that MS functionality is preserved. (b) Effect of $CO_2$ concentration on $CO_2$ desorption profiles for 0, 400, and 3000 ppm of $CO_2$ in $N_2$ following the same heat (red) + hydration (blue) desorption protocol used in panel (a).

Regarding water desorption, IRA900-RA releases more $H_2O$ than Lewatit (Figure 5c), with most of the desorption occurring near 100 °C. In contrast, Lewatit desorbs water at lower temperatures. Kinetic analysis reveals that both materials exhibit two rate constants for $H_2O$ desorption (Figure 5d): a faster initial step, followed by a slower step. For lRA900-RA, the initial rate constant is 0.06 $min^{-1}$, which is an order of magnitude slower than that of Lewatit at 0.36 $min^{-1}$. The secondary step in IRA900-RA proceeds at 0.006 $min^{-1}$, an order of magnitude slower, while in Lewatit, the second step is negligible, with a near-zero rate constant. Again, this illustrates that the binding energy of $H_2O$, in addition to the binding energy of $CO_2$, is greater in IRA900-RA over that of Lewatit.

To evaluate $CO_2$ and $H_2O$ desorption in IRA900-RA, we applied the hybrid desorption protocols described in Section 2.3.1, combining heat and hydration steps. The resulting $CO_2$ desorption profiles are shown in Figure 6a. In both sequences, $CO_2$ desorption was observed under both heating and hydration steps, confirming the responsiveness of IRA900-RA to both thermal and moisture-driven regeneration. For comparison, the same heat + hydration protocol was applied to Lewatit and no additional $CO_2$ release was observed during the hydration step. This confirms that the moisture-driven $CO_2$ release is a distinctive feature of the reactive anion chemistry in IRA900-RA, not seen with state-of-the-art amine-based sorbents. The difference in final $CO_2$ capacity depending on the order of steps (heat-first vs hydration-first) likely arises from polymer decomposition when hydration is applied first. The formation of hydroxide from bicarbonate decomposition destabilizes the $QA^+$ groups and accelerates backbone degradation, leading to additional $CO_2$ release.

To better understand the rate of $CO_2$ desorption under each condition, we analyzed the $CO_2$ desorption data using the same pseudo-first-order model (eq 10), as shown in Figure 6b and c. Low-temperature hydration showed faster desorption kinetics compared to thermal desorption regardless of the order of these steps. This illustrates the unique feature of the chemisorption MS, where $CO_2$ binding energetics to an anion are shifted through changing hydration states. Additionally, the extracted mass transfer coefficients (k) were not significantly affected by the order in which the heat and hydration steps were applied. It is noted that the faster moisture-driven desorption of IRA900 is still an order of magnitude slower than the amine thermal desorption (Figure 5b).

To assess whether these desorption protocols affect the $CO_2$ capacity of IRA900-RA, we measured $CO_2$ uptake at 400 ppm of $CO_2$ under 0% RH for 5 h. As shown in Figure 7a, the hydration-first protocol resulted in less $CO_2$ re-sorption (~20% of the desorbed $CO_2$) compared to the heating-first protocol (~50% resorption). Based on these results, the heat-first $CO_2$ desorption protocol was chosen for the water sorption experiments on IRA900. This approach allowed thermal desorption to first remove $H_2O$ to a base case as Lewatit. The thermal desorption step also minimizes variability in initial water content and better isolates the effect of the subsequent hydration step. Beginning the experiment from a controlled dry state ensures that the observed moisture-driven $CO_2$ release could be more directly attributed to the resin's reactive anion chemistry rather than differences in baseline moisture. Furthermore, the MS responsiveness of IRA900-RA was preserved after the heat-first desorption protocol, as demonstrated by $CO_2$ desorption during exposure to 400 ppm of $CO_2$ at 95% RH for 5 h.

*3.2.2. $CO_2$ Desorption under Mixed $N_2$/$CO_2$.* To investigate $CO_2$ desorption in anion exchange sorbents under more realistic sorbent regeneration conditions, we performed additional desorption experiments on IRA900-RA using the same heat + hydration protocol, but in the presence of either 400 or 3000 ppm of $CO_2$ in $N_2$, simulating $CO_2$-enriched environments.

Figure 7b shows that the amount of $CO_2$ desorbed during thermal desorption decreases as the background $CO_2$ concentration increases. This trend is consistent with Le

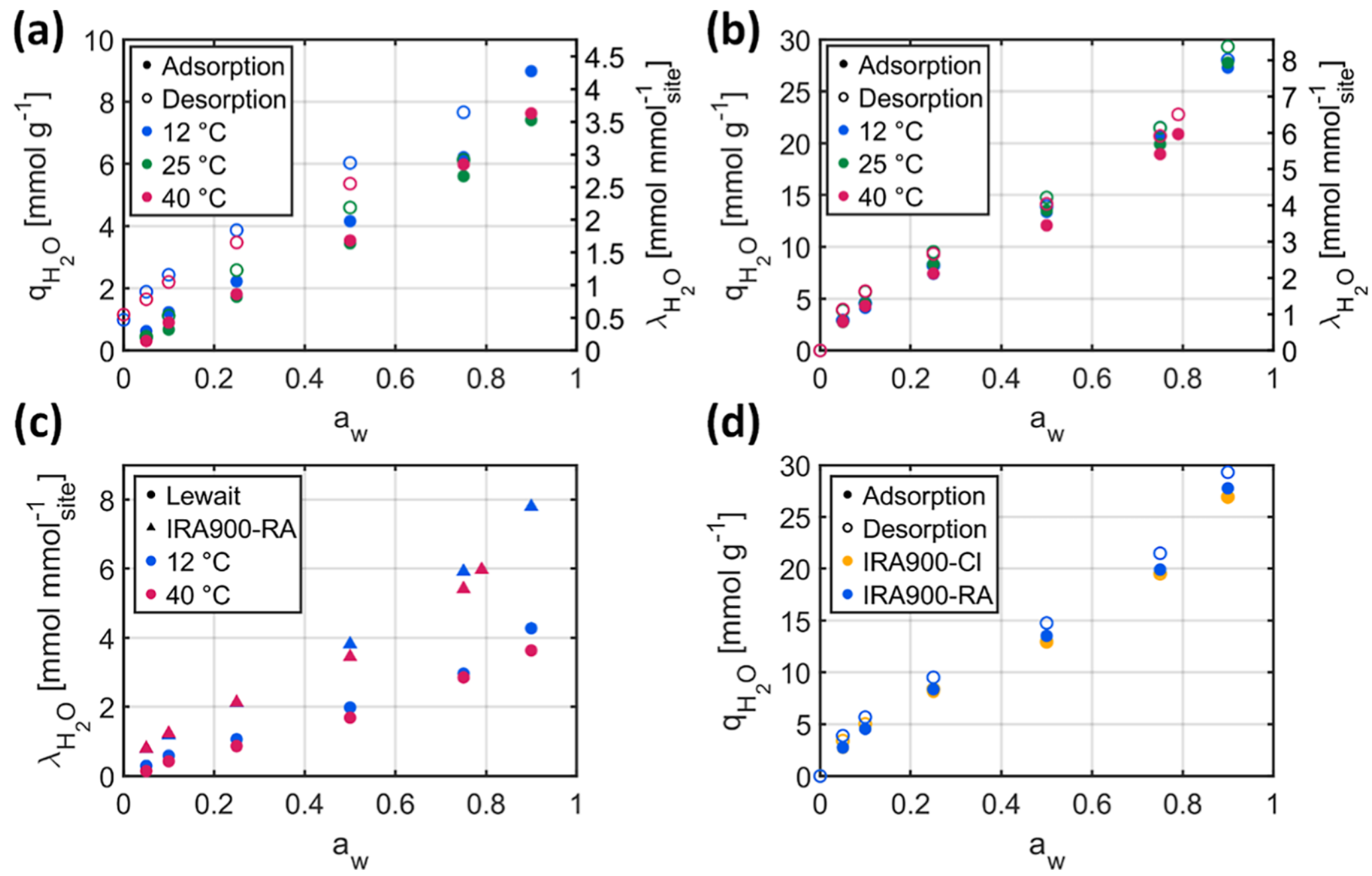

**Figure 8.** Water vapor sorption isotherms of (a) Lewatit and (b) IRA900-RA at 12 °C, 25 °C, and 40 °C, showing cumulative water uptake $q_{H_2O}$ (left $y$-axis) and site-normalized uptake $\lambda_{H_2O}$ (right $y$-axis) as functions of RH. (c) Overlay of the sorption branches of $\lambda_{H_2O}$ for Lewatit and IRA900-RA at 12 and 40 °C. (d) Comparison of IRA900-RA and IRA900-Cl isotherms at 25 °C to investigate the effect of the anion form on water sorption and hysteresis.

Chatelier's principle, which predicts that higher partial pressures of $CO_2$ will shift the equilibrium toward the sorbed state, reducing desorption (eq 5). At 3000 ppm of $CO_2$, the measured $CO_2$ release during thermal desorption falls within the uncertainty range of the gas analyzer. A similar trend is observed during the hydration-driven desorption step, where the presence of elevated $CO_2$ suppresses $CO_2$ desorption, again consistent with Le Chatelier's principle. The release of $CO_2$ with hydration at 3000 ppm of $CO_2$ illustrates the increased gas-phase concentration that can be achieved in the separation by simply adjusting water activity. Note, this is not a competitive sorption effect, as Lewatit does not exhibit a similar behavior (Figure 6a).

**3.3. Water Sorption in $N_2$.** Water sorption isotherms were obtained from TGA measurements to quantify cumulative water uptake on a dry polymer weight basis, $q_{H_2O}$ [mmol g$^{-1}$], as a function of RH in Lewatit and IRA900-based sorbents. Figures 8a and b show $q_{H_2O}$ as a function of RH during sorption and desorption at 12 °C, 25 °C, and 40 °C for Lewatit and IRA900-RA, respectively. To facilitate interpretation, a secondary $y$-axis shows $\lambda_{H_2O}$ [mmol mmol$^{-1}{}_{sites}$], calculated by normalizing $q_{H_2O}$ by the respective functional group site densities (Section 2.1). Therefore, the $\lambda$ axis reflects the number of water molecules associated with each amine or reactive anion site and provides a more direct comparison of site-normalized hydration between the materials. The $\lambda_{H_2O}$ values for each RH level are listed in Table S2 (Lewatit) and Table S3 (IRA900-RA).

Both materials roughly exhibit Type II isotherms, noted by the slight concave down shape relative to water activity at low relative humidity, followed by a nearly linear region at mid RH levels and ending with a steeper increase in sorption at higher humidities.[33] However, in Lewatit, the isotherm shape shows a temperature-induced transition. Lewatit's isotherm at 12 °C exhibits a more pronounced knee at the completion of monolayer coverage, which is a characteristic feature of Type II behavior, suggesting stronger sorbent−sorbate interactions and distinct monolayer formation. At 25 and 40 °C, this feature becomes subtle, indicating a shift toward Type III behavior. This observation is consistent with IUPAC definitions, which attribute Type III isotherms to relatively weak sorbent−sorbate interactions. Notably, Young et al.[16] previously reported Type III behavior for Lewatit at temperatures higher than 25 °C. Our data suggests that Lewatit may undergo a temperature-induced transition from Type II to Type III behavior. In contrast, IRA900-RA maintains a Type II isotherm shape across all temperatures with a persistent knee point. This difference suggests that water uptake in IRA900-RA is governed by strong, electrostatic interactions between water dipoles and permanent charges, while Lewatit's uptake is governed by weaker hydrogen-bonding interactions.

Type II behavior has been seen for gas and vapor sorption in glassy polymers,[34,35] like the glassy structures encountered in these lightly cross-linked polystyrene structures. Type II and III are further seen with physisorption of vapor and gases on macroporous surfaces where the sorbate can form multiple sorption layers,[14,16,33] as expected in the hydration clouds of water surrounding an anion. Given the exothermic nature of the sorption, water uptake is expected to decrease slightly over the 12−40 °C range. Overall, the temperature dependence of the isotherms on a water activity basis is small. In contrast, the differences due to hysteresis between the sorption and desorption branches are more pronounced.

Hysteresis between sorption and desorption branches was observed in both IRA900-RA and Lewatit across all temperatures, consistent with known behavior in porous polymeric materials and glassy polymers.[14,16,52,53] Hysteresis is commonly attributed to a combination of factors, including changes in free volume and sorption capacity, swelling-induced increase in sorption site accessibility that persists during desorption due to slow structural relaxation, kinetic limitations during desorption, and true thermodynamic equilibrium effects associated with the formation of stable species. While Lewatit generally exhibited greater hysteresis, likely due to its smaller pores and higher surface area, IRA900-RA showed a more temperature-sensitive hysteresis trend.

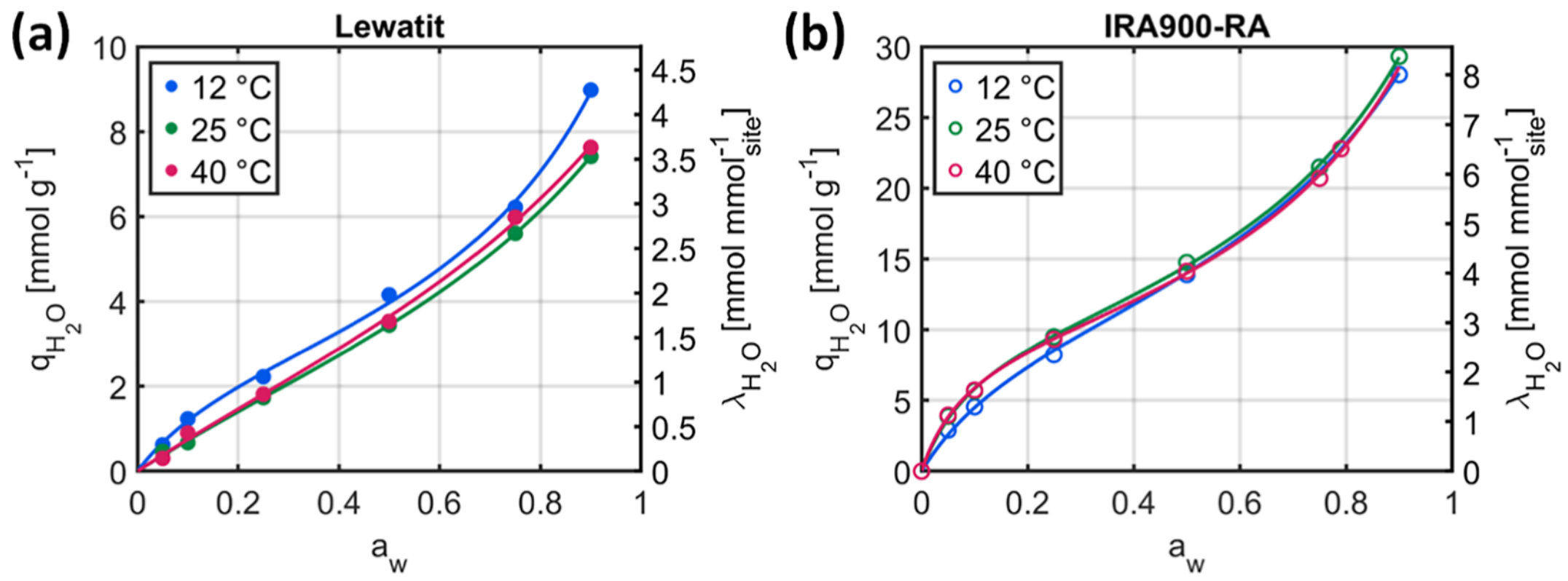

**Figure 9.** Water vapor isotherms at 12 °C, 25 °C, and 40 °C for (a) Lewatit (sorption branches) and (b) IRA900-RA (desorption branches), with GAB model fitting demonstrating good agreement with the experimental data.

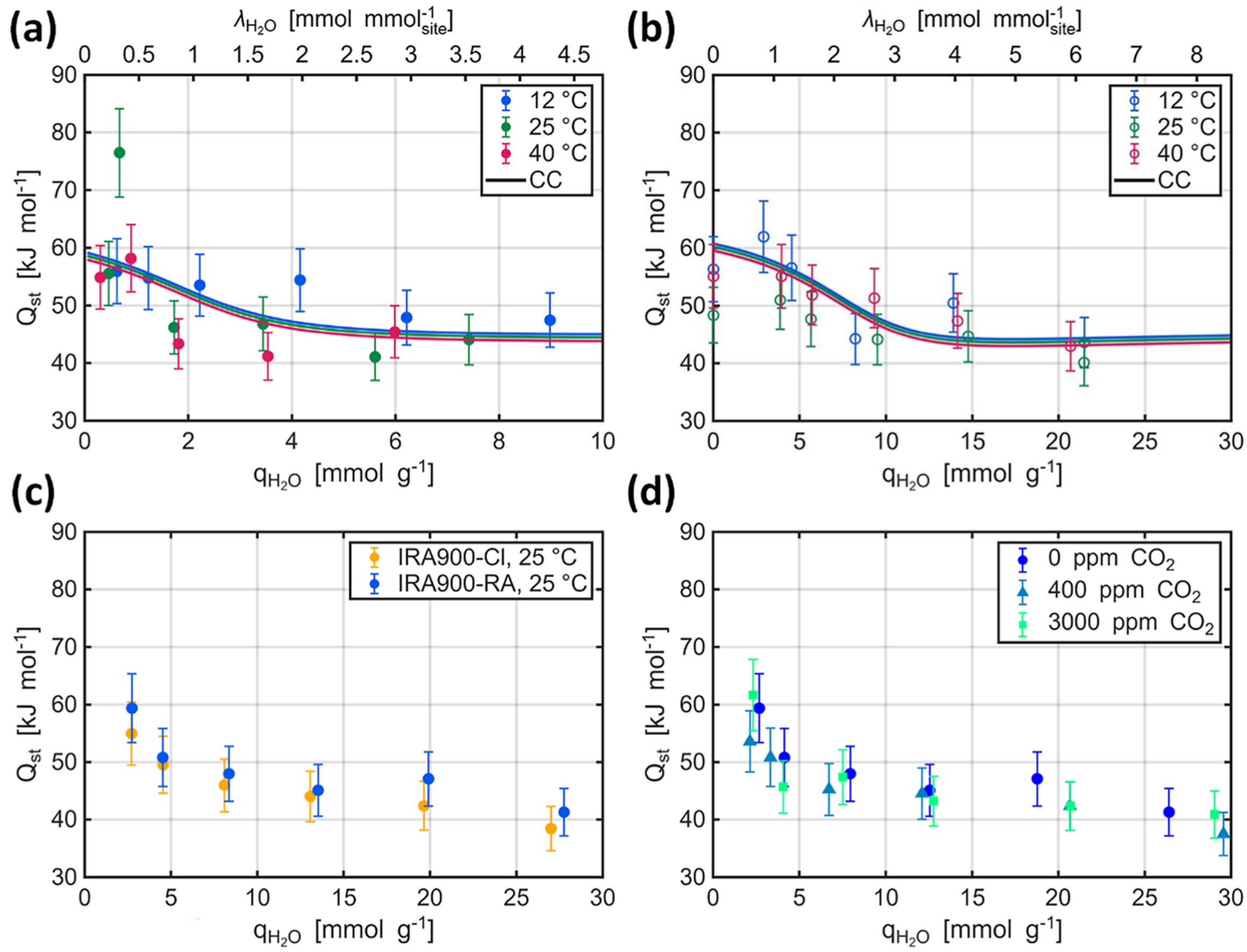

**Figure 10.** Molar heat of water sorption (|$Q_{st}$|) as a function of water uptake ($q_{H_2O}$) obtained from direct calorimetric measurement (scatter points) and Clausius−Clapeyron (CC) analysis (lines) at 12 °C, 25 °C, and 40 °C for (a) Lewatit and (b) IRA900-RA. Overlaid data points correspond to direct calorimetric measurements using STA, with error bars reflecting experimental uncertainty. (c) Comparison of $Q_{st}$ between IRA900-Cl and IRA900-RA at 25 °C, showing slightly higher heats of sorption for RA over the chloride form. (d) Water sorption enthalpies at 25 °C for IRA900-RA under pure $N_2$, 400 ppm of $CO_2$, and 3000 ppm of $CO_2$.

A direct comparison of water sorption between the two functional groups is shown in Figure 8c. This comparison highlights the higher water solubility across all RH levels in permanently charged IRA900-RA over the primary amine-based Lewatit. On an active site basis, the IRA900 loading is nearly double that of the Lewatit. Differences in water sorption capacity can be driven by differences in glassy polymer nonequilibrium void spaces, the volume of the mesoporous structure, which will be discussed based on BET surface area and pore size analysis (Figure 3 and Table S1) or governed by differences in water affinity of the distinct functional groups.

Despite its lower surface area and total pore volume (Figure 3 and Table S1), IRA900 in both Cl and RA forms exhibited higher water uptake than Lewatit across all RH values and temperatures (Figure 8a−c). The contrast between greater $N_2$ physisorption in Lewatit and greater water sorption in IRA900 may suggest that water-accessible sites emerge through polymer relaxation or hydration-specific interactions that are not accessible to $N_2$ at 77 K.

Further, since the polymer backbone structures are similar across both sorbent materials, differences in glassy non-equilibrium void space are expected to be small. Because IRA900 shows nearly twice the water uptake capacity of Lewatit on the basis of functional group density (Figure 8c), it is clear that water sorption is primarily governed by the chemical nature of the functional groups rather than the textural properties and site density. This distinction is further reflected in the GAB model parameters (Section 3.3.1). While a slight temperature dependence is observed in the isotherms of Figure 8c, it is relatively weak. The key difference lies in the stronger affinity of water for the permanently charged $QA^+$-RA ion pairs in IRA900-RA.

To investigate the influence of the anion form on water uptake and hysteresis, Figure 8d and associated data in Table S4 compare IRA900-Cl and IRA900-RA at 25 °C. The overall isotherm shapes remain the same (Type II), but the reactive anion form (likely in the $OH^-$ state) retains more total water after each desorption step, with differences ranging from 8.2−12.5%, decreasing slightly at higher RH. The slightly lower water uptake of IRA900-Cl compared to IRA900-RA may be due to differences in counterion hydration energetics between $Cl^-$ and the RA (likely $OH^-$). According to Marcus,[54] $OH^-$ (−430 kJ/mol) and $HCO_3^-$ (−335 kJ/mol) exhibit more negative hydration free energies than $Cl^-$ (−270 kJ/mol), indicating more stable water binding and stronger hydration in the RA form.

*3.3.1. Water Isotherm Fitting.* The GAB model with temperature-independent parameters $q_m$, k, and c (eq 6) was fitted to the experimental water sorption data at the three temperatures using the nonlinear least-squares regression method. The sorption branches of the $q_{H_2O}$-RH isotherms for Lewatit and the desorption branches for IRA900-RA were used for fitting. For each temperature, the fitted GAB parameters and the normalized root-mean-square error (NRMSE) are summarized in Table S5 and Table S6 for Lewatit and IRA900-RA, respectively. The experimental data and corresponding GAB fits are shown in Figure 9a and b.

The higher *c* values for IRA900-RA across all temperatures indicate that water molecules bind more strongly to primary sorption sites, consistent with stronger electrostatic interactions with the $QA^+$-RA ion pairs. The similar *k* values between IRA900-RA and Lewatit suggest that the multilayer water has comparable structure and mobility in both materials, with similar resemblance to bulk liquid water.[40] Additionally, the larger $q_m$ values in IRA900-RA reflect a greater density of energetically favorable monolayer sorption sites, further supporting the dominant role of chemical functionality over textural properties in governing sorption.

*3.3.2. Water Sorption Energetics.* Figures 10a and b show the calorimetrically measured integral heat of water sorption for Lewatit and IRA900-RA, achieved during step changes in water activity and plotted as a function of water loading, $q_{H_2O}$, with secondary axes showing $\lambda_{H_2O}$. This data correspond to the water isotherms in Figure 8 and Tables S2 and S3. Negative enthalpy values were recorded during exothermic water sorption and positive values during endothermic water desorption. For clarity, absolute enthalpy values are shown. Overlaid with the calorimetric measurements are the total isosteric sorption enthalpies estimated from the Clausius−Clapeyron approach. Calorimetric enthalpies corresponding to each $\lambda_{H_2O}$ are listed in Table S2 (Lewatit) and Table S3 (IRA900-RA). No strong temperature dependence was observed for either material.

All sorbent materials across all temperatures investigated show a decrease in enthalpy magnitude with increasing water load, eventually plateauing near the heat of water condensation. This trend reflects stronger initial binding of water molecules at low $q_{H_2O}$, where water vapor molecules primarily interact with the polymer surfaces or functional groups. If these surfaces are polar or charged, large enthalpies result due to strong electrostatic interactions of the hydrogen bond and ion−hydrogen bond. As $q_{H_2O}$ increases, additional water layer forms, transitioning from monolayer to multilayer adsorption, where interactions increasingly resemble bulk water condensation rather than surface adsorption.[16,40]

For Lewatit, Low et al.[14] obtained water sorption enthalpies via van't Hoff analysis of equilibrium isotherms and reported a range from approximately 60 kJ $mol^{-1}$ at low loadings (0−0.5 mmol $g^{-1}$) to a plateau around 40 kJ $mol^{-1}$ at higher loadings, closely matching the heat of water condensation. These values reflect a transition from monolayer resin−water interactions to multilayer water−water interactions. Additionally, Young et al.[16] used an integral water sorption enthalpy of −46 kJ $mol^{-1}$, to represent energetics of water−Lewatit interactions in their $H_2O$−$CO_2$ coadsorption model. Together, these reports support the validity of our DSC-derived values, which fall within a similar range. This supports the validity of using this approach to interpret the enthalpy of water sorption in IRA900-RA, for which, to our knowledge, no prior direct measures exist. Compared with indirect methods, DSC offers the advantage of directly measuring the heat of interaction without relying on thermodynamic assumptions. When implemented in a simultaneous TGA/DSC setup, as in this study, it also enables the efficient acquisition of enthalpy data alongside sorption isotherms, minimizing the need for separate experimental runs. However, it should be noted that in open-flow systems, convective heat loss during extended sorption processes may lead to partial underestimation of the total heat signal. Despite this, the direct calorimetric approach offers a straightforward and complementary method for assessing water−sorbent energetics.

Figure 10c and Table S7 show the comparison of the water sorption enthalpies for IRA900-RA and IRA900-Cl at 25 °C. IRA900-RA consistently exhibits slightly more exothermic enthalpies than IRA900-Cl across all water loadings. While a similar trend is also observed during desorption, the comparison is shown for sorption because the corresponding $q_{H_2O}$ at each RH step is more closely aligned between the two materials. The more exothermic sorption enthalpies in IRA900-RA suggest that the reactive anion (e.g., $OH^-$) has stronger interactions with water molecules over the chloride

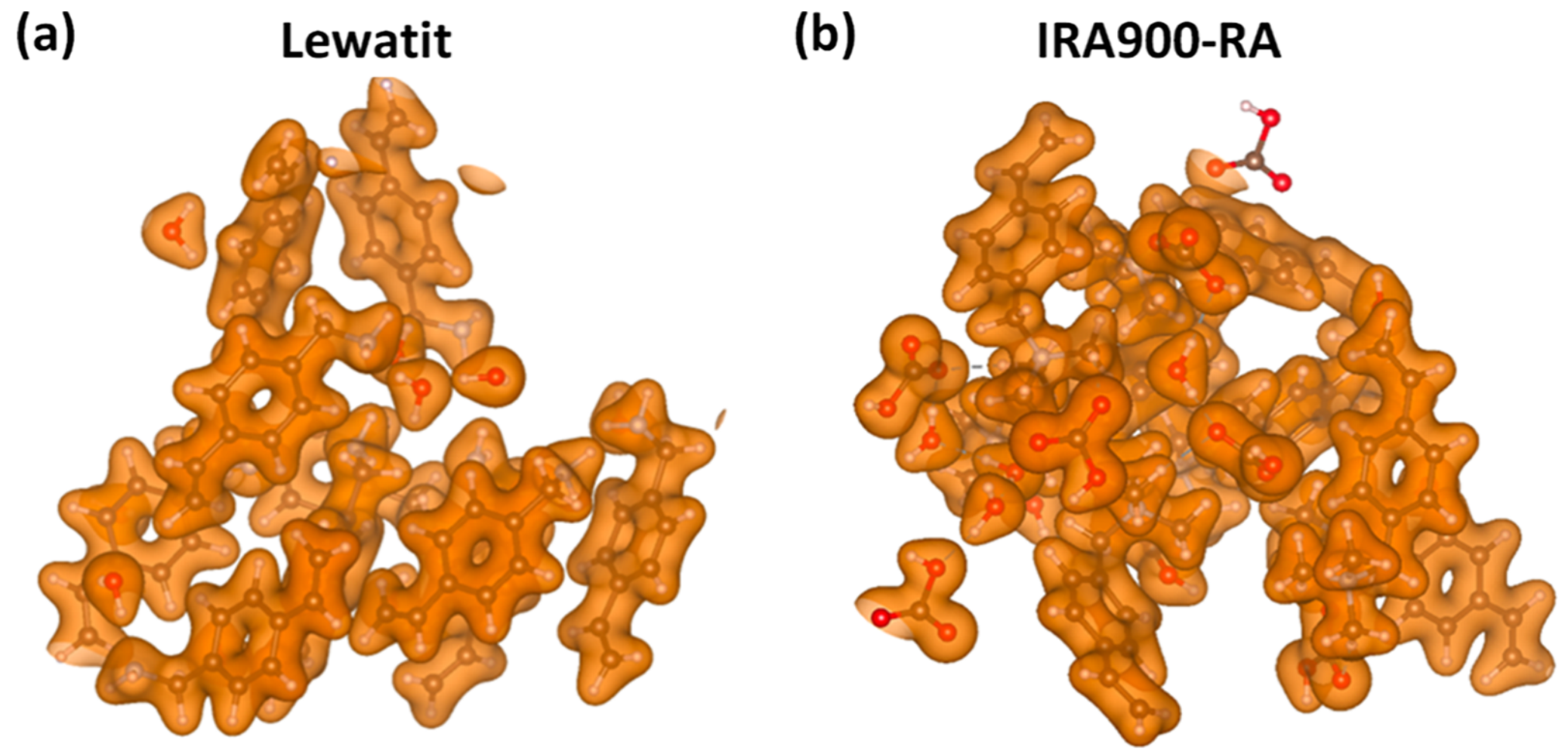

**Figure 11.** Total charge density distributions of (a) Lewatit and (b) IRA900-RA at a hydration level of $\lambda_{H_2O}$ = 1 mmol mmol$^{-1}$ site. The electron density isosurface is displayed at a consistent value of 0.05 e/$a_0^3$ ($a_0$: Bohr radius) for both structures to enable a direct comparison.

form, consistent with hydration enthalpy estimates of anions in dilute aqueous solution.

Additional calorimetric measurements were performed at 25 °C under 400 and 3000 ppm of $CO_2$ in $N_2$ to assess the effect of $CO_2$ on water sorption enthalpies for IRA900-RA (Table S8). No clear trend was observed in the enthalpy profiles as a function of the $CO_2$ concentration (Figure 10d). In these $CO_2$-containing experiments, the reported enthalpies were calculated by integrating the heat flow during each RH step and dividing by the moles of water taken up during that integration interval. The contribution of $CO_2$ to mass change was subtracted from the molar water loading calculation, but it was not possible to isolate $CO_2$-related heat contributions, specifically the endothermic heat flow from $CO_2$ desorption, that is quantified in Section 3.4. Thus, heat flow measurements in the presence of $CO_2$ are expected to be lower due to the counter-sorption of the two gases. Only slight decreases in enthalpy were observed with $CO_2$ at water loadings corresponding to humidity above 50%, but this was not outside the bounds of measurement uncertainty. This suggests that under the tested conditions, the presence of $CO_2$ does not significantly influence the heat of sorption or that any such effect is below the detection sensitivity of our open-flow calorimetric setup.

To further validate these direct calorimetry results, GAB model fits were used to calculate the isosteric heat of water sorption using the Clausius−Clapeyron (i.e., van't Hoff) expression (eq 7).[40,55−57] The enthalpies derived from the Clausius−Clapeyron equation applied to multitemperature isotherms are also overlaid onto Figure 10a and b for comparison.

Both methods capture the expected decrease in enthalpy with increasing water loading followed by a plateau near the enthalpy of water condensation (∼44 kJ/mol). The Clausius−Clapeyron method is a widely used indirect approach that complements direct calorimetry by estimating sorption enthalpies over a broader range of water loadings under equilibrium conditions. While sensitive to experimental variability in relative humidity and temperature control, it avoids the transient heat losses that can affect open-flow calorimetric setups.

Comparisons between isosteric enthalpies derived from sorption isotherms and calorimetric enthalpies have been used in sorption studies to assess whether model parameters reflect real thermodynamic interactions or structural effects.[57] In our case, the Clausius−Clapeyron-derived enthalpies based on GAB model fits agree in both trend and magnitude with directly measured calorimetric enthalpies. This agreement supports the view that the GAB fitting parameters reflect physically meaningful water−polymer interactions.

In both Lewatit and IRA900-RA, the calorimetric sorption enthalpy profile implies a local maximum at low water loadings ($q_{H_2O} \approx 1$), well below the monolayer capacity predicted by the GAB model ($q_m$), despite the GAB-derived monolayer capacity of IRA900-RA being much higher than that of Lewatit. This observation suggests that strongest, most specific sorption sites are occupied well before monolayer coverage and that these early interactions dominate the initial enthalpic response regardless of total sorption capacity. The Clausius−Clapeyron-derived enthalpy profile, in contrast, does not show a clear maximum in this range, likely due to the method's assumption of uniform, equilibrium sorption behavior and its averaging over site heterogeneity.

*3.3.3. Molecular Modeling Estimates of Water Enthalpy.* To compare the energetics of water uptake in Lewatit and IRA900-RA, we calculate the water adsorption energy, $E_{ads}$, using the case of $\lambda_{H_2O}$ = 1 mmol mmol$^{-1}{}_{site}$ as an example. Here, $E_{ads}$ is calculated using the formula: $E_{ads} = (E_{polymer} + N \cdot E_{water} - E_{polymer/water})/N$, where $E_{polymer/water}$ is the total energy of the hydrated polymer system, $E_{polymer}$ is the energy of the dry polymer in the same simulation box. $E_{water}$ is the energy of an isolated water molecule. $N$ is the number of nitrogen sites, equal to 10 for Lewatit and 8 for IRA900-RA. The calculated $E_{ads}$ values are 33.8 kJ/mol for Lewatit and 46.6 kJ/mol for IRA900-RA, which are slightly lower but in good quantitative agreement with our calorimetry measurements of ∼52 and ∼58 kJ/mol, respectively. Both theoretical and experimental results suggest that water binds more strongly to IRA900-RA than to Lewatit, indicating a higher affinity of the former for water uptake under comparable hydration levels.

Figure 11 displays the total charge density distributions of Lewatit and IRA900-RA, which provide insights into the differing electronic environments that may influence their affinities for water. In Lewatit, the charge density appears more localized and dispersed, with distinct regions of high electron density surrounding the functional group and visible gaps in between. This spatial separation suggests a less uniform electrostatic environment, that may limit the ability of the polymer to effectively stabilize water molecules. In contrast, IRA900-RA displays a denser and more interconnected charge distribution. The charge density regions are more continuous and overlap significantly with the locations of water molecules, particularly around the functional group. This implies a stronger and more cooperative electrostatic field that can better stabilize adsorbed water through hydrogen bonding and dipole interactions. These differences in charge density align with the computed adsorption energies, supporting the conclusion that IRA900-RA binds water more strongly than Lewatit due to its more favorable electronic environment.

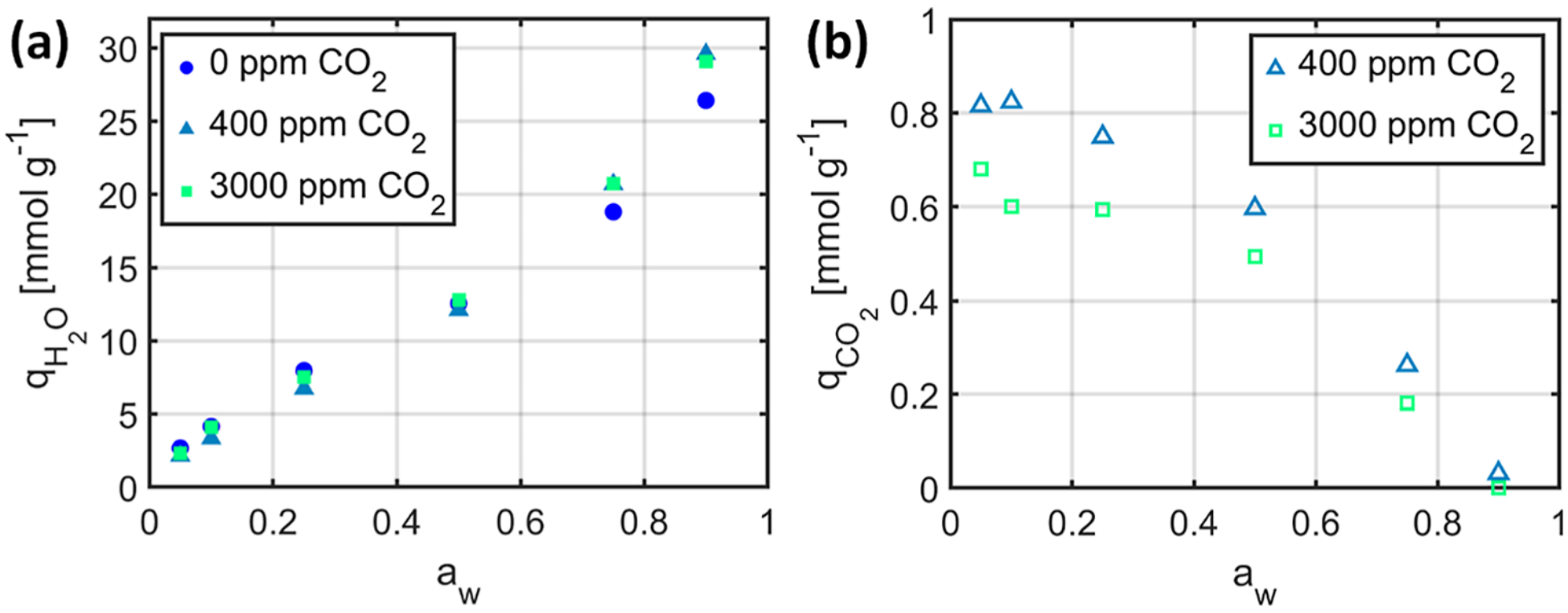

**Figure 12.** (a) Effect of $PCO_2$ (0, 400, and 3000 ppm) on water sorption in IRA900-RA at 25 °C and (b) effect of water activity on $CO_2$ loading at 400 and 3000 ppm.

**3.4. Mixed Water/$CO_2$ Counter Sorption.** To evaluate how background $CO_2$ concentration influences the water sorption behavior of IRA900-RA, additional experiments were conducted at 25 °C under three gas conditions: 0, 400, and 3000 ppm of $CO_2$ in $N_2$. These water sorption isotherms followed the heat + hydration $CO_2$ desorption protocol described in Section 3.2.2 for mixed-gas environments.

Figures 12a and S2a show the effect of $CO_2$ concentration on water loading ($q_{H_2O}$) as a function of RH during water sorption and desorption, respectively. Simultaneously, $CO_2$ desorption during water sorption and $CO_2$ sorption during water desorption ($CO_2$ and $H_2O$ counter sorption) were monitored and are shown in Figures 12b and S2b, respectively. Figures 12 and S2 illustrate the coupled dynamics of $H_2O$ and $CO_2$ interactions with the $QA^+$-based anion exchange sorbent. The numerical data for water and $CO_2$ is provided in Tables S9 and S10, respectively, and the raw data from the 400 ppm experiment is given as illustration in Figure S1.

During water (de)sorption cycles (Figures 12a and S2a), there are differences in water loading due to the presence of $CO_2$ in the system, but the sign of the effect is dependent on the equilibrated water activity. At RH <50%, the water loading is similar across all three partial pressures of $CO_2$, with an average increase of ~23% in water loading in the absence of $CO_2$ (0 ppm) compared to 400 ppm and ~8% compared to 3000 ppm. Above 50% RH, the trend reverses with higher water loading in the presence of $CO_2$, with an average increase of ~10% in water sorption at 3000 ppm compared to 0 ppm of $CO_2$. Little difference is observed between the 400 and 3000 ppm cases. The difference in water loading at low- and high-water activities is likely due to differences in hydration of the reactive anions. In the absence of $CO_2$, the original bicarbonate anions are expected to release $CO_2$ into a hydroxide state during the $CO_2$ desorption preconditioning, according to the reverse of eq 4. This was further supported by the release of $CO_2$ matching the anion exchange capacity of the IRA900 sorbent shown in Figure 6, suggesting full decomposition to the hydroxide state. Thus, the water is hydrating a $QA^+-OH^-$ ion pair in the 0 ppm of $CO_2$ experiment. In 400 ppm of $CO_2$, the reactive anion has been shown through solid-state NMR[26] and surface enhanced Raman[27] studies to exist largely in the bicarbonate state under low relative humidity and the carbonate state under high relative humidity (eq 5). At even higher $PCO_2$, the bicarbonate anion should increase according to the same chemical equilibria, in line with lower $CO_2$ desorption in Figure 12b. Thus, in the presence of $CO_2$ and at low humidity, water is hydrating predominately $QA^+$-$HCO_3^-$ ion pairs. So, at low humidity ($a_w$ <0.5), we are comparing hydration of predominant hydroxide at 0 ppm of $CO_2$ versus bicarbonate at 400 and 3000 ppm of $CO_2$. The slightly higher hydration in the 0 ppm experiment is likely due to the greater hydration number of a smaller $OH^-$ over $HCO_3^-$, which has been measured for aqueous systems.[54,58] In contrast, at high humidity ($a_w$ >0.5), the greater hydration in the $CO_2$-containing experiments is due to differences in hydrating divalent carbonate over hydroxide species. The MS mechanism predicts a higher overall level of hydration of divalent carbonates over the hydrolyzed monovalent anions ($OH^-$ or $HCO_3^-$).[10,23,59] The same $PCO_2$ dependence on water loading is less evident in the water desorption, $CO_2$ sorption cycle (Figure S2a). However, below 50% RH, the $CO_2$ free system (RA = $OH^-$) shows ~35% less water release compared to both the 400 and 3000 ppm of $CO_2$ cases, again aligning with the higher hydration requirement of the $OH^-$.

It is noteworthy that we detect differences in water sorption with $CO_2$ (Figure 12a) yet do not detect discernible differences in the corresponding enthalpy of sorption (Figure 10d) that exceed measurement uncertainty. An increase in sorption at constant temperature implies either more negative exothermic enthalpy or smaller decreases in entropy (e.g., $\Delta G = \Delta H - T\Delta S$). The small difference in enthalpy suggests more disordered water binding in the sorbent system interacting with $CO_2$. However, we note again that the enthalpy of sorption of $H_2O$ in the presence of $CO_2$ is convoluted by the counter sorption of $CO_2$, so the exact effect of $CO_2$ on water sorption enthalpy is unclear.

Finally, Figure 12b illustrates the reciprocal effect of water activity on the $CO_2$ loading ($q_{CO_2}$). As expected from the MS mechanism (eq 5), $CO_2$ is desorbed as water vapor activity increases, and vice versa (Figure S2b). The exponential drop in $CO_2$ loading with water activity is a characteristic feature of the water-dependent $CO_2$ isotherm in alkaline anion exchange resins.[9,11,31] Higher $PCO_2$ (3000 ppm) limits the reversible $CO_2$ desorption, thus leading to a reduced working capacity from the water activity change. This effect is further evident from the $CO_2$ desorption experiments illustrated in Figure 7b. Comparing the sorption of $H_2O$ to the desorption of $CO_2$ (Figure 12), we see that 30 mmol/g of water binds relative to 0.8 mmol/g of $CO_2$ released. This corresponds to a ratio of 38:1 $H_2O$:$CO_2$ on a molar basis or 15:1 on a mass basis. Although MS sorption occurs at low temperatures, it comes at the cost of water processing and partially lost through evaporation. This $H_2O$:$CO_2$ ratio is expected to improve (lower) in actual separation processes like a vapor-assisted vacuum desorption,[17,32,60] where water vapor sweeps $CO_2$ from the sorbent and leaves behind lower $PCO_2$ for further recovery. In practice, a full swing of moisture between 0 and 95% is unlikely, because ambient conditions constrain the lower bound of humidity. More efficient water separations on a $H_2O$:$CO_2$ basis are expected when the sorbent cycles between 20−50% RH when dry and near 100% RH when wet. Future work will delineate optimized sorption cycles that harness the moisture-swing effect for improved $CO_2$ recovery with substantially lower energy and water requirements.

## 4. CONCLUSION

This study compares $H_2O$ and $CO_2$ (de)sorption thermodynamics and kinetics in two classes of mesoporous polymeric chemisorbents, primary amine- and $QA^+$-based ion exchange resins, using simultaneous thermal gravimetric−calorimetric analysis, evolved gas measurements, and molecular modeling. Calorimetric water sorption enthalpies were quantified and validated against total isosteric enthalpies derived from the Clausius−Clapeyron analysis. Both approaches confirm monolayer−multilayer sorption behavior, with higher enthalpies at low water loading and enthalpies approaching the heat of water condensation at higher water loadings. The $QA^+$-based material exhibited slightly stronger monolayer water binding, supported by both enthalpy analysis and molecular modeling that shows denser charge localization and stronger $H_2O$ interaction energies at low water loadings ($\lambda_{H_2O} = 1$).

While thermal degradation limits the compatibility of the $QA^+$-based sorbent with thermal swing regeneration, low-temperature moisture exposure can induce $CO_2$ release in the QA-based material, confirming its responsiveness to moisture-driven changes in the anion hydration state, a behavior not observed in the amine sorbent.

Type II/III water sorption isotherms for both materials were described using the GAB model. Despite having a lower surface area and total pore volume, the $QA^+$-based resin showed greater water uptake per site, highlighting the dominant role of electrostatic interactions between water molecules and $QA^+$-RA pairs in governing hydration behavior.

Mixed water/$CO_2$ (de)sorption experiments revealed that not only does water influence $CO_2$ binding, but $CO_2$ also modulates water uptake through changes in anion hydration state. This coupling of water and $CO_2$ is a key feature of MS sorbents that enables energy-efficient separation of $CO_2$ under ambient conditions. The experimental and theoretical framework developed in this study to understand $H_2O$−$CO_2$ cosorption in mesoporous polymeric sorbents enables a quantitative understanding of sorbent behavior. Specifically, the measured thermodynamic parameters, including differential enthalpies of sorption and cosorption isotherms, alongside kinetic data inform process-relevant metrics such as working capacity, selectivity, energetic loads, and costs. These insights facilitate the development of efficient, scalable sorption-based separations such as direct air capture.

## ASSOCIATED CONTENT

### Supporting Information

The Supporting Information is available free of charge at https://pubs.acs.org/doi/10.1021/acsami.5c12939.

Uncertainties in mass, enthalpy, and gas composition; representative TGA, DSC, and gas analysis data for Lewatit and IRA900-RA; BET surface area and pore characteristics; water sorption isotherm data and corresponding calorimetric enthalpies across materials and temperatures; GAB model fit parameters; calorimetric enthalpies from mixed water/$CO_2$ countersorption experiments at 25 °C; mixed water desorption−$CO_2$ sorption profiles at 400 and 3000 ppm of $CO_2$; and water/$CO_2$ (de)sorption data from mixed counter-sorption experiments at 25 °C (PDF)

## AUTHOR INFORMATION

### Corresponding Author

**Jennifer L. Wade** − *The Steve Sanghi College of Engineering, Mechanical Engineering, Northern Arizona University, Flagstaff, Arizona 86011, United States;* orcid.org/0000-0001-8372-7800; Email: jennifer.wade@nau.edu

### Authors

**Golnaz Najaf Tomaraei** − *The Steve Sanghi College of Engineering, Mechanical Engineering, Northern Arizona University, Flagstaff, Arizona 86011, United States*

**Sierra Binney** − *The Steve Sanghi College of Engineering, Mechanical Engineering, Northern Arizona University, Flagstaff, Arizona 86011, United States*

**Ryan Stratton** − *The Steve Sanghi College of Engineering, Mechanical Engineering, Northern Arizona University, Flagstaff, Arizona 86011, United States*

**Houlong Zhuang** − *School for Engineering of Matter, Transport and Energy, Arizona State University, Tempe, Arizona 85287, United States;* orcid.org/0000-0001-7276-7938

Complete contact information is available at: https://pubs.acs.org/10.1021/acsami.5c12939

### Notes

The authors declare no competing financial interest.

## ACKNOWLEDGMENTS

This material is based upon work primarily supported by the U.S. Department of Energy, Office of Science, Office of Basic Energy Sciences, under Award Number DE-SC0023343. In addition, Sierra Binney acknowledges support by the National Science Foundation under Grant No. 2219247 (TGA/DSC methods development).

## ■ REFERENCES